\documentclass[reqno,a4paper,final]{amsart}
\usepackage{amsmath,amsfonts,amssymb,amsthm,amstext,eucal}
\usepackage{bbold}
\usepackage{array}
\usepackage[latin1]{inputenc}
\usepackage[notref,notcite]{showkeys}
\usepackage[all]{xy}
\newcommand{\half}{\tfrac{1}{2}}
\renewcommand{\d}{\partial}
\newcommand{\fg}{\mathfrak{g}}
\newcommand{\fh}{\mathfrak{h}}

\newcommand{\fso}{\mathfrak{so}}
\newcommand{\fsp}{\mathfrak{sp}}

\newcommand{\fsu}{\mathfrak{su}}
\newcommand{\fu}{\mathfrak{u}}
\newcommand{\ISO}{\mathrm{ISO}}
\newcommand{\SO}{\mathrm{SO}}
\newcommand{\Cl}{\mathrm{C}\ell}
\newcommand{\Spin}{\mathrm{Spin}}

\newcommand{\SU}{\mathrm{SU}}
\newcommand{\U}{\mathrm{U}}
\newcommand{\EE}{\mathbb{E}}
\newcommand{\RR}{\mathbb{R}}
\newcommand{\CC}{\mathbb{C}}

\newcommand{\ZZ}{\mathbb{Z}}

\newcommand{\eL}{\mathcal{L}}
\newcommand{\eM}{\mathcal{M}}

\DeclareMathOperator{\dvol}{dvol}
\DeclareMathOperator{\Mat}{Mat}
\newcommand{\be}{\boldsymbol{e}}
\newcommand{\bx}{\boldsymbol{x}}
\newcommand{\by}{\boldsymbol{y}}

\newcommand{\bC}{\boldsymbol{C}}

\newcommand{\TN}{\text{TN}}
\renewcommand{\1}{\mathbb{1}}
\newcommand{\MUNCH}[1]{\relax}
\allowdisplaybreaks[1]
\numberwithin{equation}{section}
\begin{document}
\dedicatory{In memory of Sonia Stanciu}
\title[Kaluza--Klein reductions of M-waves and MKK-monopoles]%
{Supersymmetric Kaluza--Klein reductions of M-waves and MKK-monopoles}
\author[Figueroa-O'Farrill]{José Figueroa-O'Farrill}
\address{School of Mathematics, The University of Edinburgh, Scotland,
  United Kingdom}
\email{j.m.figueroa@ed.ac.uk}
\author[Simón]{Joan Simón}
\address{The Weizmann Institute of Physical Sciences, Department of
  Particle Physics, Rehovot, Israel}
\email{jsimon@weizmann.ac.il}
\thanks{EMPG-02-14, WIS/37/02-AUG-DPP}
\begin{abstract}
  We investigate the Kaluza--Klein reductions to ten dimensions of the
  purely gravitational half-BPS M-theory backgrounds: the M-wave and
  the Kaluza--Klein monopole.  We determine the moduli space of smooth
  (supersymmetric) Kaluza--Klein reductions by classifying the
  freely-acting spacelike Killing vectors which preserve some Killing
  spinor.  As a consequence we find a wealth of new supersymmetric IIA
  configurations involving composite and/or bound-state configurations
  of waves, D0 and D6-branes, Kaluza--Klein monopoles in type IIA and
  flux/nullbranes, and some other new configurations.  Some new
  features raised by the geometry of the Taub--NUT space are
  discussed, namely the existence of reductions with no continuous
  moduli.  We also propose an interpretation of the flux 5-brane in
  terms of the local description (close to the branes) of a bound
  state of D6-branes and ten-dimensional Kaluza--Klein monopoles.
\end{abstract}

\maketitle

\tableofcontents

\section{Introduction and conclusions}
\label{sec:introconc}

In this paper we investigate and classify the possible supersymmetric
Kaluza--Klein reductions of the purely gravitational half-BPS M-theory
backgrounds to ten dimensions.  It is thus an extension of the ideas
and techniques developed in \cite{FigSimFlat} and \cite{FigSimBranes}
and applied to eleven-dimensional Minkowski spacetime and the
elementary half-BPS M2- and M5-branes, respectively.

A purely gravitational supersymmetric M-theory background $(M,g)$
consists of an eleven-dimensional lorentzian spin Ricci-flat manifold
admitting parallel spinors.  We remark, as discussed for example in
\cite{JMWaves}, that in lorentzian geometry the existence of parallel
spinors does not imply Ricci-flatness.  The existence of parallel
spinors does however constraint the holonomy to belong to the subgroup
$\Spin(7) \ltimes \RR^9$ of the eleven-dimensional Lorentz group
\cite{JMWaves, Bryant-ricciflat}.  There are precisely two half-BPS
possibilities: either a gravitational wave \cite{MWave} (which can be
delocalised along one or more transverse directions) or the product
$\RR^{1,6} \times N$ where $N$ is a four-dimensional hyperkähler
manifold.  For $N$ the Taub--NUT gravitational instanton, this is the
eleven-dimensional Kaluza--Klein monopole \cite{SMKK,GPMKK,HKMKK}.

The standard Kaluza--Klein reductions of these spaces give rise to
well-known IIA backgrounds: wave, D0-brane, Kaluza--Klein monopole and
D6-brane and, just as for flat space and the M2- and M5-branes, considering
more general (twisted) reductions one obtains backgrounds where
fluxbranes and nullbranes have been added.  Indeed, one of our main
aims is to give a complete list of the supersymmetric composite
configurations made of waves, D0-branes, D6-branes, Kaluza--Klein
monopoles in type IIA with fluxbranes and nullbranes.

The techniques and methodology used in this paper are fully explained
in \cite{FigSimBranes}, to which we refer the reader for further
details, particularly the introductory section, as well as for a more
complete list of references.  Let us simply restate very briefly the
main idea.  Given a purely gravitational M-theory background $(M,g)$
with isometry group $G$, we would like to determine in the first
instance all one-parameter subgroups $\Gamma \subset G$ whose orbits
in $M$ are spacelike and such that the quotient $M/\Gamma$ is smooth.
We then wish to single out those subgroups for which the resulting IIA
background $(M/\Gamma, h,\Phi,A_1)$ is supersymmetric.  Each
one-parameter subgroup $\Gamma$ is generated by a Killing vector
$\xi$, which we identify with an element of the Lie algebra $\fg$ of
$G$.  We are free to conjugate by $G$, since conjugate elements in
$\fg$ give rise to isometric reductions, and we are also free to
rescale the Killing vector, since this corresponds to a
reparametrisation of the orbit.  Thus we are interested in the
following equivalence relation in $\fg$:
\begin{equation*}
  X  \sim t g X g^{-1} \qquad\text{where $X\in\fg$, $t\in\RR^\times$
  and $g\in G$.}
\end{equation*}
The quotient of $\fg$ by this equivalence relation defines the moduli
space of one-parameter subgroups of $G$.  Selecting from this moduli
space those subgroups for which the orbits are spacelike and the
quotient along the orbits is smooth, we arrive at the \emph{moduli
  space of smooth reductions}.  Within this space there are loci
corresponding to those reductions which are also supersymmetric.
These loci comprise the \emph{moduli space of supersymmetric
  Kaluza--Klein reductions}, and one of the main results in this paper
is the determination of this space for the M-wave and the
Kaluza--Klein monopole.  These conditions on the reduction translate
into conditions on the Killing vector used to reduce, which we fully
analyse.

As mentioned in \cite{FigSimBranes}, it is possible to stop short of
the reduction to IIA and consider new M-theory backgrounds obtained by
quotienting by a (discrete) cyclic subgroup $\Gamma_0 \subset \Gamma$.
In the case where $\Gamma$ is noncompact (i.e., diffeomorphic to
$\RR$) one has that $\Gamma_0$ is infinite cyclic and hence isomorphic
to $\ZZ$, whereas for $\Gamma$ a circle subgroup, which will occur in
the Kaluza--Klein monopole, $\Gamma_0$ will be isomorphic to $\ZZ_N$
for some $N$.  Quotienting by $\Gamma_0$ thus gives rise to M-theory
backgrounds which are locally isometric to the original background.
Although we do not emphasise these constructions in this paper, let us
point out that from our results there also follows more or less
straightforwardly a classification of such ``orbifolds'' where the
group $\Gamma_0$ has one generator in the image of the exponential
map.

Having determined the supersymmetric Kaluza--Klein reductions, we then
use the techniques explained in detail in \cite{FigSimBranes} to pass
to adapted coordinates and write down the corresponding IIA background
explicitly in order to further identify in terms of composites or
bound states of D-branes, Kaluza--Klein monopoles, waves, fluxbranes
or nullbranes.  Briefly, we change coordinates from, say, $(z,\by)$,
where $\xi=\d_z + \alpha$, $\alpha$ standing for an arbitrary
element of $\fg$ commuting with $\d_z$, to an adapted coordinate
system $(z,\bx)$ defined by
\begin{equation}
  \label{eq:adapted}
  \bx = U \by \quad , \quad U =\exp(-z\alpha) \quad
  \text{such that} \quad \xi(z,\bx) = \d_z~.
\end{equation}
Since $\alpha$ acts affinely, the reduction manifestly depends on a
constant matrix $B$ and a constant vector $\bC$ defined by
\begin{equation}
 \label{eq:data}
 \alpha\by = B\by + \bC ~.
\end{equation}
It is then straightforward to perform the actual reductions, as
explained in detail in \cite{FigSimBranes}.

Even though it is not emphasised in this work, it should be clear that
by applying the usual dualities, one could construct a wealth of new
supersymmetric configurations involving duals of fluxbranes and
nullbranes \cite{FigSimFlat} and duals of standard waves, D0-branes,
Kaluza--Klein monopoles and D6-branes.

Let us now summarise the main results of this paper.  As already
mentioned we classify the supersymmetric configurations in type IIA
supergravity involving waves, Kaluza--Klein monopoles, D0-branes,
D6-branes, fluxbranes and nullbranes.  These results are summarised in
Tables~\ref{tab:Mwave}, \ref{tab:Mwave1}, \ref{tab:Mwave2},
\ref{tab:kk1}, \ref{tab:kk2} and \ref{tab:D6}.  As already stressed in
\cite{FigSimBranes} for the M2 and M5-branes, we find new backgrounds,
not only associated with bound states of waves and D0-branes or
monopoles and D6-branes in the flux/nullbrane sectors of the theory,
but also in the case of the delocalised M-wave, other backgrounds with
a more elusive interpretation, obtained by reducing along the orbits
of Killing vectors which involve time/lightlike translations and
transverse rotations.

The analysis of the supersymmetric reductions of the Kaluza--Klein
monopole reveals some interesting features.  In the first place there
exist reductions with only discrete moduli.  Due to the fact that the
Kaluza--Klein monopole has freely-acting Killing vectors with compact
orbits, there are further requirements to the ones discussed in
\cite{FigSimBranes} to be satisfied to get a smooth spin manifold. In
particular, the integral curves of these Killing vectors need to be
periodic and this fact manifests itself in the integrality of the
parameters defining the Killing vector.  Supersymmetry then imposes
further linear Diophantine equations on these parameters, resulting in
discrete regimes in moduli space.  Second, there exist fluxbranes
constructed out of the 3-spheres that foliate the Taub--NUT geometry.
The action of Killing vectors on the Killing spinors of the Taub--NUT
space (see Section~\ref{sec:action}) reveals the possibility of
constructing supersymmetric fluxbranes by performing Kaluza--Klein
reductions along the orbits of a Killing vector involving not only a
spacelike translation along the monopole, but also both a rotation on
the monopole and an element of $\SU(2)$ acting naturally on the
3-spheres foliating the Taub--NUT geometry. Finally, we give a novel
interpretation of the flux 5-brane \cite{GSflux} as the local
description (close to the branes) of a bound state of IIA
Kaluza--Klein monopoles and D6-branes\cite{CGBoundStates}.  This will
be argued in terms of the supersymmetry preserved by both systems, and
furthermore, by explicitly studying the supergravity configuration
describing the bound state close to the branes.

The paper is organised as follows.  In Section~\ref{sec:MWave}, we
apply our technology both to the M-wave and its delocalisation along
one transverse direction.  In Section~\ref{sec:MKK} the same is done
for the Kaluza--Klein monopole.  Some technical points concerning
group theory and spinors, Killing spinors of the Taub--NUT geometry
and the action of the isometry group on them are left to the
corresponding appendices.  Finally, an appendix is included where we
analyse the question of the existence of spin structures on quotients
of the Kaluza--Klein monopole.

\section{Kaluza--Klein reductions of the M-wave}
\label{sec:MWave}

In this section we classify the set of IIA backgrounds obtained by
reducing the M-wave along the orbits of a one-parameter subgroup of
the isometry group.  In Section~\ref{sec:MW} we discuss the M-wave and
in Section~\ref{sec:MWd} we discuss the M-wave which has been
delocalised along one transverse direction.

\subsection{M-wave}
\label{sec:MW}

In this section we discuss the supersymmetric Kaluza--Klein reductions
of the purely gravitational M-wave \cite{MWave}
\begin{equation}
  \label{eq:mMW}
  g = 2 dy^+ dy^- + 2V (dy^+)^2 + ds^2(\EE^9)~,
\end{equation}
where $V$ is a harmonic function on $\EE^9$.  The maximally symmetric
solution corresponds to a function $V$ which only depends on the
transverse radius $r$ given by $r^2 = \sum_iy^i y^i$.  Demanding that
the spacetime be asymptotically flat at large $r$ means that
$\lim_{r\to\infty} V(r)$ should be a constant.  This constant can be
reabsorbed by a change of variables, whence a convenient choice of $V$
which makes the metric manifestly asymptotically flat is
\begin{equation*}
  V(r) = \frac{Q}{r^7}~,\quad\text{for some $Q>0$.}
\end{equation*}
In the absence of $F_4$, the Killing spinors are parallel relative to
the spin connection.  In the above coordinates, such spinors are
constant
\begin{equation}
  \label{eq:SMW}
  \varepsilon = \varepsilon_\infty
\end{equation}
and obey
\begin{equation}
  \label{eq:PMW}
  \Gamma_+ \varepsilon_\infty = 0~.
\end{equation}

The isometry group is
\begin{equation}
  \label{eq:GMW}
  G = \SO(9) \times \RR^2 \subset \ISO(1,10)~,
\end{equation}
where $\RR^2$ corresponds to translations along the lightcone
directions $y^\pm$, and the $\SO(9)$ is the transverse rotation
group.  The Lie algebra is given by
\begin{equation}
  \label{eq:gMW}
  \fg = \fso(9) \times \RR^2~,
\end{equation}
whence any Killing vector can be decomposed as
\begin{equation}
  \label{eq:KVMW}
  \xi = \tau + \rho~,
\end{equation}
with $\tau = a\d_+ + b \d_-$ and $\rho \in \fso(9)$.

\subsubsection{Freely-acting spacelike isometries}

By conjugating with $G$, we may bring $\xi$ to a normal form.
In practice this means conjugating $\rho$ to belong to a fixed Cartan
subalgebra.  For example, we can choose
\begin{equation}
  \label{eq:CartanMW}
  \rho = \theta_1 R_{12} + \theta_2 R_{34} + \theta_3 R_{56} +
  \theta_4 R_{78}~,
\end{equation}
where $R_{ij}$ stands for the generator of a rotation in the
($ij$)-plane.  The norm of a Killing vector $\xi = a\d_+ + b\d_- +
\rho$ is given by
\begin{equation*}
  \|\xi\|^2 = 2 ab + 2V a^2 + \|\rho\|^2~,
\end{equation*}
which is bounded below by the flat norm of the translation component:
\begin{equation*}
  \|\xi\|^2 \geq 2 ab = \|\tau\|^2_\infty~
\end{equation*}
That this bound is sharp can be shown by simply looking at large $r$
along any direction fixed by $\rho$: $y^9$ in the above example.
Therefore $\xi$ is spacelike if and only if $\tau$ is spacelike
relative to the flat metric.  It is more convenient to change
coordinates from lightcone $(y^\pm)$ to pseudo-euclidean
$(y^0,y^\natural)$ such that $\tau = c \d_0 + d \d_\natural$.  The
condition that $\tau$ be asymptotically spacelike says that $d^2 >
c^2$, whence $d$ cannot be zero.  Using the freedom to rescale we can
set $d=1$, leaving a one parameter family of translations $\tau =
\d_\natural + c\d_0$ with $-1<c<1$.  The action generated by such a
Killing vector is always free, and hence the moduli space of smooth
reductions of the M-wave is five-dimensional and is parametrised by
the cartesian product of the interval $(-1,1)$ with a fixed Cartan
subalgebra of $\fso(9)$.  As we will see below, supersymmetry will
select a locus with codimension one.

\subsubsection{Absence of closed causal curves}

We would like to prove that the M-wave background \eqref{eq:mMW}
reduced along the orbits of $\xi=\partial_\natural + a\partial_0$
$(|a|<1)$ has no closed causal curves.  As already stressed in the
previous subsection, the constraint $|a|<1$ comes from demanding
that $\xi$ be everywhere spacelike.  The norm of $\xi$ is given by
\begin{equation*}
  \|\xi\|^2 = (1-a)\left[(1+a) + (1-a) V\right]~.
\end{equation*}
In adapted coordinates to the action of $\xi$, the background
\eqref{eq:mMW} takes the form
\begin{equation*}
  g = (V-1)(dt)^2 + \|\xi\|^2 (dz)^2 -2(1-a) V dz\cdot dt +
  ds^2(\EE^9)~.
\end{equation*}
Suppose, for a contradiction, that a causal curve $x(\lambda)$ does
exist joining the points $(t_0,z_0,x^i_0)$ and
$(t_0,z_0+\Delta,x^i_0)$ for $i = 1,2,\dots,9$.  It follows that there
must exist at least one value $\lambda^*$ of the affine parameter
where the timelike component of the tangent vector to the causal curve
must vanish.  Computing the norm of the tangent vector at that point,
one derives the inequality
\begin{equation*}
    \|\xi\|^2 (\lambda^*) \,
    \left.\frac{dz}{d\lambda}\right|^2_{\lambda^*} + \sum_i
  \left.\frac{dx^i}{d\lambda}\right|^2_{\lambda^*} \leq 0~,
\end{equation*}
which can never be satisfied unless the tangent vector to the causal
curve vanishes identically at $\lambda^*$, violating the
hypothesis that $\lambda$ is an affine parameter.  Therefore we
conclude there are no such closed causal curves.

\subsubsection{Supersymmetry}

As usual the translation component of $\xi$ is not constrained by
supersymmetry, but the rotation component is constrained to lie in the
isotropy of a parallel spinor obeying \eqref{eq:PMW}.  In
\cite{FigSimBranes} we explained how to determine the supersymmetric
locus in the parameter space and we refer to that paper for details.
Let $S_{11}$ denote the irreducible spinor representation of
$\Spin(1,10)$ and let $S_{11}^+ = \ker \Gamma_+ \subset S_{11}$ denote
the space of Killing spinors of the M-wave.  Under the transverse spin
group $\Spin(9)$, $S_{11}^+$ is isomorphic to the unique irreducible
spinor representation $S_9$.  As discussed in
Appendix~\ref{sec:groups}, relative to the basis dual to the $R_{ij}$
in \eqref{eq:CartanMW}, the subspace $S_{11}^+$ has weights
$(\pm1,\pm1,\pm1,\pm1)$ where the signs are uncorrelated, for a total
of $16$ weights.  Therefore $\rho$ will annihilate a Killing spinor if
and only if the $\theta_i$ belong to the union of the eight
hyperplanes
\begin{equation}
  \label{eq:hyperplanesMW}
  \sum_{i=1}^4 \mu_i \theta_i = 0~,\quad\text{with $\mu_i^2 = 1$.}
\end{equation}
(As usual there are only eight hyperplanes because $(\mu_i)$ and
$(-\mu_i)$ define the same hyperplane.)  A generic $\rho$ in one of
these hyperplanes will annihilate a two-dimensional subspace of
Killing spinors and hence the associated reduction will preserve
$\frac18$ of the supersymmetry preserved by the M-wave, or a fraction
$\nu = \frac1{16}$ of the supersymmetry of the eleven-dimensional
vacuum.  This corresponds to $\rho$ belonging to an $\fsu(4)$
subalgebra.  This is clearly a four-dimensional locus of the moduli
space of smooth reductions.

There is enhancement of supersymmetry if the rotation $\rho$ belongs
to two or more hyperplanes.  There are two kinds of pairwise
intersections: those planes where none of the $\theta_i$ vanish but
two pairs do, say $\theta_1 + \theta_2 = 0$ and $\theta_3 + \theta_4 =
0$. Such rotations belong to an $\fsp(1) \times \fsp(1)$ subalgebra
and preserve a fraction $\nu = \frac18$ of the supersymmetry.  The
other kind of pairwise intersection is when only one of the $\theta_i$
vanishes, say $\theta_1=0$ and hence $\theta_2 + \theta_3 + \theta_4 =
0$.  Such a rotation belongs to an $\fsu(3)$ subalgebra and the
reduction preserves again $\frac18$ of the supersymmetry.  These loci
are three-dimensional.  If a rotation belongs to three of these
hyperplanes, then two of the $\theta_i$ must vanish.  This means that
$\rho$ belongs to an $\fsu(2)$ subalgebra and the reduction preserves
$\frac14$ of the supersymmetry.  This locus is two-dimensional.
Finally there is one point in the intersection of all hyperplanes,
corresponding to $\rho = 0$.  This gives rise to a one-dimensional
locus of $\half$-BPS reductions.  These observations are summarised in
Table~\ref{tab:MWsusy}.

\begin{table}[h!]
  \begin{center}
    \setlength{\extrarowheight}{5pt}
    \begin{tabular}{|>{$}c<{$}|>{$}c<{$}|>{$}c<{$}|>{$}c<{$}|}
      \hline
      \multicolumn{1}{|c|}{Translation} & \text{Subalgebra} & \nu &
      \dim\\
      \hline
      \hline
      & \fsu(4) & \frac1{16} & 4\\
      \d_\natural + a \d_0 & \fsu(3) & \frac18 & 3\\
      & \fsp(1) \times \fsp(1) & \frac18 & 3\\
      -1<a<1 & \fsu(2) & \frac14 & 2\\
      & \{0\} & \frac12& 1 \\[3pt]
      \hline
    \end{tabular}
    \vspace{8pt}
    \caption{Supersymmetric reductions of the M-wave.  We indicate
      the spinor isotropy subalgebra to which the rotation belongs,
      the fraction $\nu$ of the supersymmetry preserved and the
      dimension of the corresponding stratum of the moduli space $\eM$
      of supersymmetric reductions.}
    \label{tab:MWsusy}
  \end{center}
\end{table}

We would like to stress that by restricting to (discrete) cyclic
subgroups $\Gamma_0\subset\Gamma$ generated by $\xi =
\partial_\natural + \rho$, the corresponding quotient manifolds
$\eM_{\text{wave}}/\Gamma_0$ would be describing the propagation of
M-waves in an eleven--dimensional fluxbrane.

\subsubsection{Explicit reductions}

There is a single type of reduction to be analysed for the M-wave, the
one whose Killing vector is written as $\xi = \partial_\natural +
\alpha$ where $\alpha$ stands for an infinitesimal affine
transformation consisting of a rotation in the space transverse to the
direction of propagation $z=x^\natural$ and a timelike translation
\begin{equation*}
  \begin{aligned}[m]
    \alpha &= a\partial_0 + \theta_1 (y^1\partial_2
    -y^2\partial_1) + \theta_2 (y^3\partial_4-y^4\partial_3) \\
    & {} + \theta_3 (y^5\partial_6-y^6\partial_5)
    + \theta_4 (y^7\partial_8-y^8\partial_7)~,
  \end{aligned}
\end{equation*} 
the parameter $a$ being bounded above in absolute value by $|a|<1$.

The matrix $B$ defined in \eqref{eq:data} characterising the
Kaluza--Klein reduction can be written in the basis
$\{x^0,x^1,\dots,x^8\}$ of the adapted coordinate system
\eqref{eq:adapted} as
\begin{equation}
  \label{eq:Bmatrixm5b}
  B= 
  \begin{pmatrix}
    0 & 0 &  0 & 0 & 0 & 0 & 0 & 0 & 0  \\
    0 & 0 & -\theta_1 & 0 & 0 & 0 & 0 & 0 & 0 \\
    0 & \theta_1 & 0 & 0 & 0 & 0 & 0 & 0 & 0 \\
    0 & 0 & 0 & 0 & -\theta_2 & 0 & 0 & 0 & 0 \\
    0 & 0 & 0 & \theta_2 & 0 & 0 & 0 & 0 & 0 \\
    0 & 0 & 0 & 0 & 0 & 0 & -\theta_3 & 0 & 0 \\
    0 & 0 & 0 & 0 & 0 & \theta_3 & 0 & 0 & 0 \\
    0 & 0 & 0 & 0 & 0 & 0 & 0 & 0 & -\theta_4 \\
    0 & 0 & 0 & 0 & 0 & 0 & 0 & \theta_4 & 0 \\
  \end{pmatrix}~.
\end{equation}
Notice that $x^9$ is left invariant by the construction giving rise to
these configurations. Since there is an extra translation operator
besides $\partial_\natural$, there is a 9-vector $\bC$ taking care of
the inhomogeneous part in the change of coordinates
\eqref{eq:adapted}.  The transpose of this vector is given by
\begin{equation*}
  (\bC)^t = (a\,,\vec{0}) ~.
\end{equation*}

After the Kaluza--Klein reduction, one obtains the ten-dimensional
metric
\begin{equation*}
  g = \Lambda^{1/2}\left\{\left(V-1\right)(dx^0)^2 +
    ds^2(\EE^9)\right\}
  - \Lambda^{3/2} A_1^2
\end{equation*}
where $A_1$ stands for the RR 1-form, which takes the form
\begin{equation*}
    A_1 = \Lambda^{-1}\left\{dx^0\left[V(1+a) - a\right] 
      + \left(B\cdot x\right)^i\cdot dx_i\right\}
\end{equation*}
in terms of a scalar function $\Lambda$ given by
\begin{equation}
   \label{eq:lwave}
   \Lambda = (1+a)\left[(1-a) + V(1+a)\right] + \left(B\cdot
     x\right)^i
   \left(B\cdot x\right)_i~.
\end{equation}
The dilaton is also given in terms of $\Lambda$ by $\Phi =
\frac{3}{4}\log\Lambda$.

Let us, first of all, discuss the physical interpretation for the
configurations described in the subspace $a=0$. It should be clear at
this stage, that these configurations describe composite
configurations of D0-branes and fluxbranes. The absence of null
rotations is telling us that there are no D0-branes in the nullbrane
sector of string theory.  For arbitrary values of the deformation
parameters $\{\theta_i\}$, the configuration would break supersymmetry
completely, and its interpretation would be in terms of composite
configurations involving D0-branes at $r=0$ and, generically, four
different F7-branes lying at $x^1=x^2=0$, $x^3=x^4=0$, $x^5=x^6=0$ and
$x^7=x^8=0$, respectively.  It is the presence of the F7-branes that
breaks supersymmetry completely.

\begin{table}[h!]
  \begin{center}
    \setlength{\extrarowheight}{3pt}
    \begin{tabular}{|>{$}c<{$}|c|>{$}l<{$}|}
      \hline
      \nu & Object & \multicolumn{1}{c|}{Subalgebra}\\
      \hline
      \hline
      \frac14 & D0$+$F5 & \fsu(2)\\
      \frac18 & D0$+$F3 & \fsu(3)\\
      \frac18 & D0$+$F1 & \fsp(1)\times \fsp(1)\\
      \frac{1}{16} & D0$+$F1 & \fsu(4)\\[3pt]
      \hline
    \end{tabular}
    \vspace{8pt}
    \caption{Supersymmetric configurations of D0-branes and
      fluxbranes}
    \label{tab:Mwave}
  \end{center}
\end{table}

On the other hand, there are four different types of supersymmetric
configurations which are summarised in Table~\ref{tab:Mwave}.  The
discussion of the taxonomy is entirely analogous to the one given for
M2-branes in \cite{FigSimBranes} when restricting to the above
subclass of Killing vectors. We refer the reader to the corresponding
subsection for further details.  Notice, though, that in this case the
D0-branes are not constrained to move on the fluxbranes, but
everywhere on $\EE^9$, as already emphasized in \cite{JoanD0F5}.

Let us finally move to the subspace where $a\neq 0$. Let us first
consider the background in which all $\theta_i$ parameters are set to
zero. In that particular case, the ten dimensional configuration turns
out to be
\begin{equation}
 \label{eq:newd0}  
  \begin{aligned}[m]
    g &= -\Lambda^{-1/2} (dx^0)^2 + \Lambda^{1/2}ds^2(\EE^9) \\
    F_2 &= dA_1 = dx^0\wedge d\Lambda^{-1} \\
    \Phi &= \tfrac34 \log\Lambda~,
  \end{aligned}
\end{equation}
where the scalar function $\Lambda$ in \eqref{eq:lwave} reduces to
\begin{equation*}
  \Lambda = (1+a)\left[(1-a) + V\,(1+a)\right]~.
\end{equation*}

Such a configuration has the same group of isometries as the standard
solution describing D0-branes. Since $|a|<1$, it is still
asymptotically flat, but its RR charge acquires an extra $(1+a)^2$
constant factor.  Due to the fact that $a$ is a continuous parameter,
whenever $a\neq 0$, the charge will no longer be quantised. As
expected, for $a=\pm 1$, the configuration becomes singular. The
physical status for these configurations remains unclear to us. It is
clear, though, that by switching the parameters $\theta_i$ in a
supersymmetric preserving way, we are adding fluxbranes to the basic
configuration \eqref{eq:newd0}.

\subsection{Delocalised M-wave}
\label{sec:MWd}

In this section we discuss the supersymmetric Kaluza--Klein reductions
of the purely gravitational M-wave when it has been delocalised in one
transverse direction.  This solution has metric
\begin{equation}
  \label{eq:mMWd}
  g = 2 dy^+ dy^- + 2V (dy^+)^2 + (dy)^2 + ds^2(\EE^8)~,
\end{equation}
where $V$ is a harmonic function on $\EE^8$ and $y$ stands for the
transverse spacelike direction in which \eqref{eq:mMW} was
delocalised.  The maximally symmetric solution corresponds to a
function $V$ which only depends on the transverse radius $r$ given by
$r^2 = \sum_i y^i y^i$.  Again a convenient choice with the virtue of
yielding a manifestly asymptotically flat spacetime is
\begin{equation*}
  V(r) = \frac{Q}{r^6}~,\quad\text{for some $Q>0$.}
\end{equation*}
The properties of its Killing spinors are exactly as in \eqref{eq:SMW}
and \eqref{eq:PMW}.

The isometry group is
\begin{equation}
  \label{eq:GMWd}
  G = \SO(8) \times \RR^3 \subset \ISO(1,10)~,
\end{equation}
where $\RR^3$ corresponds to translations along the lightcone
directions $y^\pm$ and $y$, and the $\SO(8)$ is the transverse
rotation group.  The Lie algebra is given by
\begin{equation}
  \label{eq:gMWd}
  \fg = \fso(8) \times \RR^3~,
\end{equation}
whence any Killing vector can be decomposed as
\begin{equation}
  \label{eq:KVMWd}
  \xi = \tau + \rho~,
\end{equation}
with $\tau = a\d_+ + b \d_- + c\d_y$ and $\rho \in \fso(8)$. 

\subsubsection{Freely-acting spacelike isometries}

As usual we may bring $\xi$ to a normal form by conjugating with $G$.
Notice that $\rho$ can always be chosen as in \eqref{eq:CartanMW}.
In this way, the norm of the most general Killing vector is given by
\begin{equation*}
  \|\xi\|^2 = 2 ab + 2V a^2 + c^2 + \|\rho\|^2 ~.
\end{equation*}
In contrast with the case of the M-wave treated above, the rotation
now need not leave any direction invariant.  This means that the norm
of the rotation is bounded below by $r^2 m^2$, where $m^2$ is the
minimum of the norm of $\rho$ on the unit sphere in $\EE^8$, and as a
result the norm of $\xi$ is bounded below by
\begin{equation*}
  \|\xi\|^2 \geq \|\tau\|^2_\infty + 2V(r) a^2 + r^2 m^2~.
\end{equation*}
If $\rho$ is given by \eqref{eq:CartanMW}, then $m^2 = \min_i
\theta_i^2$.  This bound is sharp since there are points in the sphere
for which $\|\rho\|^2 = r^2 m^2$.  We must therefore distinguish
between two cases:
\begin{enumerate}
\item[(a)] $a=0$.  In this case $\|\xi\|^2 \geq c^2 + r^2 m^2$, whence
  it is bounded below by $c^2$ as $r$ tends to $0$.  Therefore the
  Killing vector is everywhere spacelike if and only if $c\neq 0$.
\item[(b)] $a\neq 0$. In this case the norm of $\xi$ is bounded below
  by a function
  \begin{equation*}
    f(r) = \|\tau\|^2_\infty + \frac{2Qa^2}{r^6} + r^2 m^2~.
  \end{equation*}
  This function is bounded below.  Ensuring that the lower bound is
  positive will impose a lower bound on $\|\tau\|^2_\infty$ allowing
  it to be negative, as was already noticed for the M2-brane and the
  delocalised M5-brane in \cite{FigSimBranes}.  The function $f(r)$
  grows without bound as $r\to0$ and $r\to\infty$.  There is a unique
  critical point $r_0>0$ given by
  \begin{equation*}
    f'(r_0) = 0 \implies m^2 r_0^8 = 6 Q a^2~.
  \end{equation*}
  Demanding that $f(r_0) > 0$ yields a lower bound on the
  asymptotic norm of the translation: $\|\tau\|^2_\infty > - \mu^2$,
  where
  \begin{equation*}
    \mu^2 = \frac{8}{6^{3/4}} Q^{1/4} |a|^{1/2} m^{3/2}~.
  \end{equation*}
\end{enumerate}
In all cases the action is free provided that the translation
component is present.

In summary, we can distinguish between different kinds of
freely-acting spacelike Killing vectors:
\begin{enumerate}
\item[(A)] $\xi = \d_y + b \d_- + \rho$, where we have already used
  the freedom to rescale and put $c=1$, whence $b$ and $\rho$ are
  unconstrained.  Such Killing vectors form a five-dimensional stratum
  of the moduli space of smooth reductions.
\item[(B)] In this case $\xi = \d_+ + b\d_- + c \d_y + \rho$, where
  we have used the freedom to rescale and the fact that $a\neq 0$ to
  set $a=1$.  We must distinguish between two cases:
  \begin{enumerate}
  \item[(i)] $\rho$ does not fix any direction; whence $2b + c^2 > -
    \mu^2$ with
    \begin{equation}
      \label{eq:musqMWd}
      \mu^2 = \frac{8}{6^{3/4}} Q^{1/4} m^{3/2}~.
    \end{equation}
    Such Killing vectors give rise to a six-dimensional stratum of the
    moduli space of smooth reductions.
  \item[(ii)] $\rho$ fixes some direction, so that one of the $\theta$
    parameters in $\rho$ vanishes.  In this case, $2b + c^2 > 0$, so
    that the translation is spacelike relative to the flat norm.  The
    corresponding stratum of the moduli space is five-dimensional.
  \end{enumerate}
\end{enumerate}
In all cases, supersymmetry will select a codimension-one locus.

\subsubsection{Supersymmetry}

The analysis of supersymmetry is entirely analogous to the fully
localised M-wave configuration, since all new possibilities discussed
previously only involve translational isometries, which do not
constrain supersymmetry.  The supersymmetric reductions are summarised
in Table~\ref{tab:MWdsusy}.

\begin{table}[h!]
  \begin{center}
    \setlength{\extrarowheight}{5pt}
    \begin{tabular}{|>{$}c<{$}|>{$}c<{$}|>{$}c<{$}|>{$}c<{$}|}
      \hline
      \multicolumn{1}{|c|}{Translation} & \text{Subalgebra} & \nu &
      \dim\\
      \hline
      \hline
      & \fsu(4) & \frac1{16} & 4 \\
      & \fsu(3) & \frac18 & 3 \\
      \d_y + b\d_- & \fsp(1)\times\fsp(1) & \frac18 & 3 \\
      & \fsu(2) & \frac14 & 2 \\
      & \{0\} & \frac12 & 1 \\[3pt]
      \hline
      \d_+ + b \d_- + c \d_y & \fsu(4) & \frac1{16} & 5 \\
      2b + c^2 > - \mu^2 & \fsp(1)\times\fsp(1) & \frac18 & 4 \\[3pt]
      \hline
      \d_+ + b \d_- + c \d_y & \fsu(3) & \frac18 & 4 \\
      & \fsu(2) & \frac14 & 3 \\
       2b + c^2 > 0 & \{0\} & \frac12 & 2 \\[3pt]
      \hline
    \end{tabular}
    \vspace{8pt}
    \caption{Supersymmetric reductions of the delocalised M-wave.  We
      indicate the form of the translation, the spinor isotropy
      subalgebra to which the rotation belongs, the fraction $\nu$ of
      the supersymmetry preserved and the dimension of the
      corresponding stratum of the moduli space $\eM$ of
      supersymmetric reductions.  The parameter $\mu^2$ is nonzero and
      is given in \eqref{eq:musqMWd}.}
    \label{tab:MWdsusy}
  \end{center}
\end{table}

\subsubsection{Explicit reductions}

Let us start our explicit reduction analysis by considering the
reductions along the orbits of the Killing vector $\xi = \partial_y +
\alpha$, where $\alpha$ stands for the infinitesimal affine
transformation
\begin{multline*}
    \alpha = b\partial_- + \theta_1 (y^1\partial_2 -y^2\partial_1) 
    + \theta_2 (y^3\partial_4-y^4\partial_3) \\
    + \theta_3 (y^5\partial_6-y^6\partial_5)
    + \theta_4 (y^7\partial_8-y^8\partial_7)~.
  \end{multline*}
  
  The constant matrix $B$ defined in \eqref{eq:data} is a $9\times 9$
  matrix which in the adapted coordinate system \eqref{eq:adapted}
  does not act either on the $x^+$ or $y$ directions. It is given
  explicitly, in the basis $\{x^-,x^1,\dots,x^8\}$, by
\begin{equation}
  \label{eq:Bmatrixwave1}
  B= 
  \begin{pmatrix}
    0 & 0 & 0 & 0 & 0 & 0 & 0 & 0 & 0 \\
    0 & 0 & -\theta_1 & 0 & 0 & 0 & 0 & 0 & 0 \\
    0 & \theta_1 & 0 & 0 & 0 & 0 & 0 & 0 & 0 \\
    0 & 0 & 0 & 0 & -\theta_2 & 0 & 0 & 0 & 0 \\
    0 & 0 & 0 & \theta_2 & 0 & 0 & 0 & 0 & 0 \\
    0 & 0 & 0 & 0 & 0 & 0 & -\theta_3 & 0 & 0 \\
    0 & 0 & 0 & 0 & 0 & \theta_3 & 0 & 0 & 0 \\
    0 & 0 & 0 & 0 & 0 & 0 & 0 & 0 & -\theta_4 \\
    0 & 0 & 0 & 0 & 0 & 0 & 0 & \theta_4 & 0 \\
  \end{pmatrix}~.
\end{equation}
There is a nontrivial 9-vector $\bC$ taking care of the inhomogeneous
part of the change of coordinates due to the existence of the extra
translation $\partial_-$.  In the same basis used above, it is given
by
\begin{equation*}
  (\bC)^t = (a,\vec{0})~.
\end{equation*}

The ten dimensional configuration obtained by Kaluza--Klein reduction
has a metric that takes the form
\begin{equation*}
  g = \Lambda^{1/2}\left\{2dx^+dx^- + 2V (dx^-)^2 +
    ds^2(\EE^8)\right\} - \Lambda^{3/2} A_1^2~,
\end{equation*}
where $A_1$ is the RR 1-form potential, which together with the
nontrivial dilaton profile are given by
\begin{equation*}
  \begin{aligned}[m]
    A_1 &= \Lambda^{-1}\left(bdx^+ + (B\cdot x)^i dx_i\right) \\
    \Phi &= \tfrac34\log\Lambda
  \end{aligned}
\end{equation*}
in terms of the scalar function
\begin{equation*}
  \Lambda = 1 +  (B\cdot x)^i (B\cdot x)_i~.
\end{equation*}

If we restrict ourselves to the subspace defined by $b=0$, and set all
the rotation parameters $\theta_i$ to zero, one gets the standard wave
solution in type IIA.  It is then clear that by keeping $b=0$ but
turning on some of the $\theta_i$, one would start generating new
solutions in the fluxbrane sector.  The classification and
interpretation of the solutions is analogous to the ones found before
and are summarised in Table~\ref{tab:Mwave1}.

\begin{table}[h!]
  \begin{center}
    \setlength{\extrarowheight}{3pt}
    \begin{tabular}{|>{$}c<{$}|c|>{$}l<{$}|}
      \hline
      \nu & Object & \multicolumn{1}{c|}{Subalgebra}\\
      \hline
      \hline
      \frac14 & WA$+$F5 & \fsu(2)\\
      \frac18 & WA$+$F3 & \fsu(3)\\
      \frac18 & WA$+$F1 & \fsp(1)\times \fsp(1)\\
      \frac{1}{16} & WA$+$F1 & \fsu(4)\\[3pt]
      \hline
    \end{tabular}
    \vspace{8pt}
    \caption{Supersymmetric configurations of type IIA waves (WA) and
    fluxbranes}
    \label{tab:Mwave1}
  \end{center}
\end{table}

On the other hand, we can study the family of solutions characterised
by vanishing $\theta_i$, but having a nonvanishing extra translation
parameter $(b\neq 0)$.  Notice that the scalar function becomes
trivial $(\Lambda=1)$ and the RR 1-form potential $A_1=bdx^+$ becomes
pure gauge.  We are thus left with a purely gravitational
configuration with a wave metric
\begin{equation*}
  g = 2dx^+dx^- + (2V-b^2)(dx^+)^2 + ds^2(\EE^8)~.
\end{equation*}
Notice that this spacetime is again asymptotically flat in the limit
$r\to\infty$, since in this limit $2V - b^2$ tends to a constant.
Turning on the angle parameters $\theta_i$ one is just adding
fluxbranes to the above configuration.

There is a second inequivalent set of Kaluza--Klein reductions that
one can study for these backgrounds. These are the reductions along
the orbits of the Killing vector $\xi=\partial_+ + \alpha$, where
$\alpha$ stands for the generators of transverse rotations in $\EE^8$
and translations in the $\{x^-\,,y\}$ directions :
\begin{multline*}
  \alpha = b\partial_- + c\partial_y +
  \theta_1 (y^1\partial_2 -y^2\partial_1) 
  + \theta_2 (y^3\partial_4-y^4\partial_3) \\
  + \theta_3 (y^5\partial_6-y^6\partial_5)
  + \theta_4 (y^7\partial_8-y^8\partial_7)~.
\end{multline*}

The constant matrix $B$ defined in \eqref{eq:data} is a $10\times 10$
matrix.  It is given explicitly, in the adapted coordinate
\eqref{eq:adapted} basis $\{x^-,y,x^1,\dots,x^8\}$, by
\begin{equation}
  \label{eq:Bmatrixwave2}
  B= 
  \begin{pmatrix}
    0 & 0 & 0 & 0 & 0 & 0 & 0 & 0 & 0 & 0 \\
    0 & 0 & 0 & 0 & 0 & 0 & 0 & 0 & 0 & 0 \\
    0 & 0 & 0 & -\theta_1 & 0 & 0 & 0 & 0 & 0 & 0 \\
    0 & 0 & \theta_1 & 0 & 0 & 0 & 0 & 0 & 0 & 0 \\
    0 & 0 & 0 & 0 & 0 & -\theta_2 & 0 & 0 & 0 & 0 \\
    0 & 0 & 0 & 0 & \theta_2 & 0 & 0 & 0 & 0 & 0 \\
    0 & 0 & 0 & 0 & 0 & 0 & 0 & -\theta_3 & 0 & 0 \\
    0 & 0 & 0 & 0 & 0 & 0 & \theta_3 & 0 & 0 & 0 \\
    0 & 0 & 0 & 0 & 0 & 0 & 0 & 0 & 0 & -\theta_4 \\
    0 & 0 & 0 & 0 & 0 & 0 & 0 & 0 & \theta_4 & 0 \\
  \end{pmatrix}~.
\end{equation}
There is a nontrivial 10-vector $\bC$ taking care of the inhomogeneous
part of the change of coordinates \eqref{eq:adapted} due to the
existence of the extra translations $\partial_+$ and $\partial_y$. In
the same basis used above, it is given by
\begin{equation*}
  (\bC)^t = (a,c,\vec{0})~.
\end{equation*}

After reduction, the ten-dimensional metric becomes
\begin{equation*}
  g = \Lambda^{1/2}\left\{ds^2(\EE^8) + (dx)^2\right\} -
  \Lambda^{3/2}(A_1)^2 ~,
\end{equation*}
where $dy=dx + c dx^+$ and $A_1$ is the RR 1-form potential given,
together with the dilaton, by
\begin{equation*}
  \begin{aligned}[m]
    A_1 &= \Lambda^{-1}\left(dx^- + c\,dx + (B\cdot x)^i dx_i\right)
    \\
    \Phi &= \tfrac34\log\Lambda~,
  \end{aligned}
\end{equation*}
where
\begin{equation*}
  \Lambda = 2b + c^2 + 2V(r) +  (B\cdot x)^i (B\cdot x)_i~.
\end{equation*}

As indicated in Table~\ref{tab:MWdsusy}, whenever $\theta_i\neq 0$ for
all $i$, the two extra translation parameters must satisfy the bound
$2b+c^2>-\mu^2$, where $\mu$ is given in \eqref{eq:musqMWd}.  We do
not have a physical understanding for this configuration.

On the other hand, when $\theta_i=0$ for all $i$, the bound is
$2b+c^2>0$. In the particular case of vanishing $c$, the configuration
looks like a delocalised D0-brane, in which $x^-$ is playing the role
of a timelike coordinate after the reduction.  Thus, whenever $c\neq
0$, and following similar arguments to the ones presented in
\cite{FigSimBranes} when dealing with similar reductions, one could
interpret the corresponding background as a bound state of type IIA
waves and delocalised D0-branes. Thus, by turning on different
$\theta$ parameters, we are adding fluxbranes to these bound states,
and thus breaking further supersymmetry. The possible supersymmetric
configurations are summarised in Table~\ref{tab:Mwave2}.

\begin{table}[h!]
  \begin{center}
    \setlength{\extrarowheight}{3pt}
    \begin{tabular}{|>{$}c<{$}|c|>{$}l<{$}|}
      \hline
      \nu & Object & \multicolumn{1}{c|}{Subalgebra}\\
      \hline
      \hline
      \frac14 & (WA+D0)$+$F5 & \fsu(2)\\
      \frac18 & (WA+D0)$+$F3 & \fsu(3)\\
      \frac18 & (WA+D0)$+$F1 & \fsp(1)\times \fsp(1)\\
      \frac{1}{16} & (WA+D0)$+$F1 & \fsu(4)\\[3pt]
      \hline
    \end{tabular}
    \vspace{8pt}
    \caption{Supersymmetric configurations of bound states made of
      type IIA waves and delocalised D0-branes (WA+D0) and fluxbranes}
    \label{tab:Mwave2}
  \end{center}
\end{table}

\section{Supersymmetric reductions of the Kaluza--Klein monopole}
\label{sec:MKK}

In this section we classify the supersymmetric reductions of the
eleven-dimensional Kaluza--Klein monopole: a half-BPS purely
gravitational M-theory background isometric to the product of the
seven-dimensional Minkowski spacetime with a noncompact
four-dimensional hyperkähler space: the Taub--NUT space.

The metric is given by \cite{SMKK,GPMKK}
\begin{equation}
  \label{eq:mMKK}
  g = ds^2(\EE^{1,6}) + g_{\TN}~,
\end{equation}
where $g_{\TN}$ is the Taub--NUT metric to be described below.
The isometry group is
\begin{equation}
  \label{eq:GMKK}
  G = \ISO(1,6) \times \U(2) \subset \ISO(1,6) \times \SO(4) \subset
  \ISO(1,10)~,
\end{equation}
with Lie algebra
\begin{equation}
  \label{eq:gMKK}
  \fg = \left( \RR^{1,6} \rtimes \fso(1,6) \right) \times \fsu(2)
  \times \fu(1)~.
\end{equation}

Since $F_4$ also vanishes for this background, the Killing spinors are
the parallel spinors relative to the spin connection, just as for the
M-waves discussed in the previous section.  Since the background is
metrically a product and one factor is flat, the parallel
(complexified) spinors are given by the tensor products $\eta \otimes
\varepsilon$, where $\eta$ is a parallel spinor of Minkowski spacetime
and $\varepsilon$ is a parallel spinor in Taub--NUT.  As shown in
Appendix~\ref{sec:PSTN}, $\varepsilon$ is given by equation
\eqref{eq:PSTN} where $\varepsilon$ obeys \eqref{eq:chiral} and hence
has positive chirality.  Counting dimensions we see that the
Kaluza--Klein monopole is indeed a half-BPS background \cite{HKMKK}.

\subsection{The Taub--NUT geometry}

We shall next briefly discuss the geometry of the Taub--NUT space.
(For a review see, e.g., \cite{EGH}.) The Taub--NUT metric is given by
\begin{equation}
  \label{eq:mTN}
  g_{\TN} = V ds^2(\EE^3) + V^{-1} (d\chi + A)^2~,
\end{equation}
where the function $V:\EE^3 \to \RR$ and the gauge field
$A$ are related by the abelian monopole equation
\begin{equation*}
  F_A := dA = -\star dV~,
\end{equation*}
where $\star$ is the Hodge star in $\EE^3$.  It follows from this
equation that $V$ is harmonic and that $F_A$ obeys Maxwell's equations
$d\star F_A = 0$.  Although it is possible to consider multicentred
solutions, we will consider for simplicity the maximally
symmetric case
\begin{equation*}
  V(r) = 1 + \frac{Q}{r}~,\quad\text{for some $Q>0$},
\end{equation*}
where $r$ is the euclidean radius in $\EE^3$.  The corresponding
solution of Maxwell's equations is the Dirac monopole.  In spherical
polar coordinates $(r,\theta,\varphi)$ for $\EE^3$, where the metric
is given by
\begin{equation*}
  ds^2(\EE^3) = dr^2 + r^2 d\theta^2 + r^2 \sin^2\theta d\varphi^2
\end{equation*}
and the orientation by
\begin{equation*}
  \dvol(\EE^3) = r^2 \sin\theta dr \wedge d\theta \wedge d\varphi~,
\end{equation*}
the gauge field can be chosen to be
\begin{equation}
  \label{eq:ATN}
  A = -Q \cos\theta d\varphi~.
\end{equation}
The field-strength is proportional to the volume form on the unit
sphere in $\EE^3$:
\begin{equation*}
  F_A = Q \sin\theta d\theta \wedge d\varphi~,
\end{equation*}
whence the charge of monopole is given by
\begin{equation*}
 \frac{1}{4\pi} \int_{S^2} F_A = Q~.
\end{equation*}

The Taub--NUT metric is therefore defined on the total space of the
circle bundle over $\EE^3$ (minus the origin) corresponding to a Dirac
monopole of charge $Q$ at the origin.  Restricted to the unit sphere
(and hence to any sphere) in $\EE^3$, this circle bundle is the Hopf
fibration with total space a $3$-sphere.  These $3$-spheres are the
orbits under the isometry group $\U(2) \subset \SO(4)$ of the
Taub--NUT metric, whence we see that it acts with cohomogeneity one.
More precisely, the orbits are parametrised by $r\geq 0$.  The generic
orbits, which occur for $r>0$, are $3$-spheres, and there is a
degenerate orbit at $r=0$ consisting of a point: the \emph{nut}, with
due apologies to Newman, Unti and Tamburino.  The foliation of the
Taub--NUT space (minus the nut) by $3$-spheres is analogous to the
foliation of $\RR^4$ (minus the origin) by $3$-spheres: the difference
is that whereas the spheres in $\RR^4$ are round with isometry group
$\SO(4)$, the spheres in Taub--NUT are squashed with isometry group
$\U(2) \subset \SO(4)$.  A manifestation of this fact is that
asymptotically (as $r\to\infty$),
\begin{equation*}
  g_{\TN} \sim dr^2 + r^2 d\theta^2 + r^2 \sin^2\theta d\varphi^2 +
  (d\chi - Q \cos\theta d\varphi)^2~,
\end{equation*}
whence the Taub--NUT metric has a circle which remains of constant
size, hence squashing the sphere, instead of growing as it would have
to in order to keep the sphere round.

We can rewrite the Taub--NUT metric so that the isometries are
manifest.  This is made easier by identifying the orbits with the Lie
group $\SU(2)$ on which we have a natural action of $\SU(2) \times
\SU(2)$ by left and right translations.  The centre acts trivially,
whence we have an action of the quotient $\SO(4)$ which restricts to
the action of the subgroup $\U(2)$.  Let
\begin{equation}
  \label{eq:sigmas}
  \begin{split}
    \sigma_1 &= -\cos\psi d\theta + \sin\theta\sin\psi d\varphi\\
    \sigma_2 &= -\sin\psi d\theta - \sin\theta\cos\psi d\varphi\\
    \sigma_3 &= d\psi - \cos\theta d\varphi~,
  \end{split}
\end{equation}
denote the right-invariant Maurer--Cartan forms in the Lie group
$\SU(2)$.  The range of the angular coordinates are $0\leq \theta \leq
\pi$, $\varphi \in \RR/2\pi\ZZ$ and $\psi \in \RR/4\pi\ZZ$.  One
checks that
\begin{equation}
  \label{eq:struceqn}
  d\sigma_1 = \sigma_2 \wedge \sigma_3~, \qquad
  d\sigma_2 = \sigma_3 \wedge \sigma_1 \qquad\text{and}\qquad
  d\sigma_3 = \sigma_1 \wedge \sigma_2~.
\end{equation}

Identifying $\chi$ with $Q\psi$, we can rewrite the Taub--NUT metric
\eqref{eq:mTN} as
\begin{equation}
  \label{eq:metricTN}
  g_{\TN} = V dr^2 + Vr^2 (\sigma_1^2 + \sigma_2^2) + V^{-1} Q^2
  \sigma_3^2~.
\end{equation}
Because it is written using the Maurer--Cartan forms, the invariance
under $\SU(2)$ is manifest.  There is an additional $\U(1)$ symmetry
because of the fact that the coefficients of $\sigma_1^2$ and
$\sigma_2^2$ coincide.

In the limit as $r\to 0$, the Taub--NUT metric becomes
\begin{equation*}
  g_{\TN} \sim \frac{Q}{r} dr^2 + Q r (\sigma_1^2 + \sigma_2^2 +
  \sigma_3^2)~.
\end{equation*}
Changing coordinates to $\varrho = 2 \sqrt{Qr}$, we obtain a more
familiar metric
\begin{equation*}
  g_{\TN} \sim d\varrho^2 + \tfrac14 \varrho^2 (\sigma_1^2 +
  \sigma_2^2 + \sigma_3^2)~,
\end{equation*}
which is the flat metric for $\RR^4$, thought of as $\CC^2$ with
complex coordinates
\begin{equation}
  \label{eq:C2}
  \begin{split}
    z_1 = x_1 + i x_2 &= \varrho \cos(\theta/2) e^{i (\psi +
      \varphi)/2}\\
    z_2 = x_3 + i x_4 &= \varrho \sin(\theta/2) e^{i (\psi -
      \varphi)/2}~.
  \end{split}
\end{equation}
For our chosen range of coordinates $\varrho \geq 0$, $\theta \in
[0,\pi]$, $\varphi \in \RR/2\pi\ZZ$ and $\psi \in \RR/4\pi\ZZ$, the
above parametrisation covers $\RR^4$ once.  We conclude that the
Taub--NUT metric is regular (and flat) at the nut and hence can be
extended to all of $\RR^4$.

The Taub--NUT Killing vectors can be explicitly calculated as follows.
From the expression for the Taub--NUT metric \eqref{eq:mTN} it follows
that $\d_\chi$ (equivalently $\d_\psi$) is a Killing vector, in fact
it generates translations along the Hopf fibre.  The euclidean metric
in $\EE^3$ and the function $V$ are invariant under rotations in
$\EE^3$, but the gauge field $A$ is not invariant and hence they are
not isometries.  This can be easily fixed.  These rotations do leave
invariant the field-strength $F_A$, and hence they leave the gauge
field invariant up to a compensating gauge transformation. In other
words, we can modify the rotation Killing vectors in $\EE^3$ by a
suitable gauge transformation (i.e., a translation along the fibre) in
such a way that the connection one-form $d\chi + A$ is invariant.
Explicitly, let $\rho$ be a generic rotation Killing vector in
$\EE^3$.  The fact that $F_A$ is invariant means that
\begin{equation*}
  \eL_\rho dA = d \eL_\rho A = 0 \implies \eL_\rho A = d\Lambda_\rho~,
\end{equation*}
for some function $\Lambda_\rho$.  This means that $A$ is invariant up
to a gauge transformation, whence $\xi := \rho - \Lambda_\rho \d_\chi$
leaves invariant $d\chi + A$ and hence the Taub--NUT metric.  Notice
that $\Lambda_\rho$ is only defined up to a constant.  We choose this
constant in order that the Killing vectors obey the $\fsu(2) \times
\fu(1)$ algebra.  In terms of the orbit coordinates
$(\theta,\varphi,\psi)$, the Killing vectors are given by
\begin{equation}
  \label{eq:KVTN}
  \begin{split}
    \xi_1 &= -\sin\varphi\d_\theta - \cot\theta\cos\varphi\d_\varphi
    - \frac{\cos\varphi}{\sin\theta}\d_\psi \\
    \xi_2 &= -\cos\varphi\d_\theta + \cot\theta\sin\varphi\d_\varphi
    + \frac{\sin\varphi}{\sin\theta}\d_\psi\\
    \xi_3 &= \d_\varphi\\
    \xi_4 &= \d_\psi~.
  \end{split}
\end{equation}
One can also check directly that the first three such vectors leave
the $\sigma_i$ invariant, whereas the fourth rotates $\sigma_1$ and
$\sigma_2$, which provides an alternative proof that they leave the
metric invariant.  In fact, the first three are the left-invariant
vector fields on the Lie group $\SU(2)$ and, as can be easily checked,
satisfy $[\xi_i,\xi_j] = \varepsilon_{ijk} \xi_k$ for $i,j,k=1,2,3$.
The remaining vector field $\xi_4$ is right-invariant and commutes
with the other three.  In other words, these Killing vectors define a
realisation of $\fsu(2) \times \fu(1)$, with $\xi_{1,2,3}$ spanning
$\fsu(2)$ and $\xi_4$ spanning $\fu(1)$.

\subsection{Freely-acting spacelike isometries}

Our main physical motivation is to study IIA configurations involving
D6-branes and flux- and nullbranes, but to obtain the D6-brane in type
IIA we need to reduce the Kaluza--Klein monopole along the Hopf fibre
of the Taub--NUT space.  This is generated by the vector field $\xi_4$
above which vanishes at the nut.  In fact, upon reduction, the IIA
solution has a naked singularity at the nut, which is where the
D6-branes lie.  Since we are interested in generalising this reduction
in order to incorporate fluxbranes, we will allow for isometries which
are null when $r=0$, but spacelike everywhere else:
\begin{equation*}
  \|\xi\|^2\geq 0 \qquad\text{and}\qquad \|\xi\|^2 > 0 \quad\text{for
    $r>0$.}
\end{equation*}
This is analogous to allowing isometries of brane backgrounds which
are spacelike everywhere but at the brane horizon, as we did in
\cite{FigSimBranes}.  We beg the reader's indulgence in allowing us
the slight abuse of notation in referring to these Killing vectors as
\emph{spacelike} within the confines of this section.

It follows from the structure \eqref{eq:gMKK} of the Lie algebra of
isometries of the Kaluza--Klein monopole, that the most general
infinitesimal isometry can be written as
\begin{equation*}
  \xi = \tau + \lambda + \rho_{\TN}~,
\end{equation*}
where $\tau\in\RR^{1,6}$, $\lambda\in\fso(1,6)$ and $\rho_{\TN} \in
\fsu(2) \times \fu(1)$.  Notice that $\rho_{\TN}$ is orthogonal to
$\tau + \lambda$ and its norm is positive-definite except at $r=0$,
where it vanishes.  We can use the freedom to conjugate by $G$
in order to bring these Killing vectors to a normal form.  We will
treat both factors separately.

First let us consider the Taub--NUT factor.  Conjugating by $\SU(2)$
allows us to bring $\rho_{\TN}$ to the form
\begin{equation*}
  \rho_{\TN} = a \d_\psi + b \d_\varphi~,
\end{equation*}
for some constants $a,b$.  The norm of this vector field is positive
away from the nut provided that $a \pm b \neq 0$.  Indeed,
\begin{equation*}
  \|\rho_{\TN}\|^2 =   b^2 V(r) \sin^2\theta + V(r)^{-1} Q^2 \left( a
  - b \cos\theta\right)^2~,
\end{equation*}
which is clearly positive for $r>0$ unless $a = \pm b$.

The analysis of the Minkowski factor is similar to the ones given in
\cite{FigSimBranes} for the M2-brane and M5-brane backgrounds.
Conjugating by an isometry, we may bring the Lorentz component to one
of the following normal forms:
\begin{enumerate}
\item $\lambda = \theta_1 R_{12} + \theta_2 R_{34} + \theta_3 R_{56}$;
\item $\lambda = \beta B_{01} + \theta_2 R_{34} + \theta_3 R_{56}$,
  $\beta\neq 0$; or
\item $\lambda = N_{+2} + \theta_2 R_{34} + \theta_3 R_{56}$,
\end{enumerate}
where $B_{0i}$ and $N_{+i}$ stand, respectively, for an infinitesimal
boost and null rotation in the $i$th direction.  Conjugating by the
translation subgroup $\RR^{1,6}$, we can bring the translation to a
normal form depending on the form of $\lambda$.  In case (1), with
none of the $\theta_i$ vanishing, the translation can be made
proportional to $\d_0$; but then the norm of $\tau + \lambda$ would be
negative at some points, unless $\tau = 0$.  If (at least) one of the
$\theta_i$ were to vanish, say $\theta_1 = 0$, then $\tau$ can be
taken to be any translation in the $(01)$ plane.  Its norm cannot be
negative, otherwise there would be points where $\xi$ would have
negative norm: this means that $\tau$ can be either spacelike or null.
In case (2), we can take $\tau$ proportional to $\d_2$; but the boost
would make $\xi$ have negative norm at points.  This case is therefore
discarded.  Finally, in case (3), if none of the $\theta_i$ vanish,
$\tau$ can be taken to be proportional to $\d_-$, but then $\xi$ would
have negative norm at some points unless $\tau = 0$.  On the other
hand, if one of the $\theta_i$ vanish, say $\theta_2 = 0$, then we can
take $\tau$ proportional to $\d_3$.

In contrast to the M-wave and the backgrounds discussed in
\cite{FigSimBranes}, in the Kaluza--Klein monopole there are spacelike
Killing vectors which generate circle actions instead of
$\RR$-actions.  Such Killing vectors have no proper translations and
hence the analysis of whether they give rise to smooth reductions is
more delicate than in previous backgrounds.  To see what can go wrong,
simply notice that the integral curves of the vector field $a \d_\psi
+ b \d_\varphi$ away from the nut in Taub--NUT lie generically in a
torus.  If these curves fail to be periodic (that is, if the ratio
$a/b$ is not rational), the corresponding reduction would not even be
Hausdorff.  As we will see, this will manifest itself in reductions
which have no continuous moduli.

Therefore we will distinguish between two types of spacelike Killing
vectors $\xi = \tau + \lambda + \rho_{\TN}$, according to whether
$\tau$ vanishes or not.  From the observations made above, those with
nonzero $\tau$ fall into three cases:
\begin{enumerate}
\item[(A)] $\xi = \d_+ + \theta_2 R_{34} + \theta_3 R_{56} + a
  \d_\psi + b \d_\varphi$, where $a \neq \pm b$.  Such vector fields
  comprise a four-dimensional stratum of the moduli space of smooth
  reductions.
\item[(B)] $\xi = \d_1 + \theta_2 R_{34} + \theta_3 R_{56} + a \d_\psi
  + b \d_\varphi$.  The corresponding stratum is four-dimensional.
\item[(C)] $\xi = \d_3 + N_{+2} + \theta_3 R_{56} + a \d_\psi +
  b\d_\varphi$.  This stratum is three-dimensional.
\end{enumerate}
Notice that we have already used the freedom to rescale the Killing
vector.  In all of these cases, the action of $\xi$ integrates to a
free action of $\RR$.  We will see below that supersymmetry will
select a codimension-one locus.

Similarly, there are two cases with $\tau=0$:
\begin{enumerate}
\item[(a)] $\xi = \theta_1 R_{12} + \theta_2 R_{34} + \theta_3 R_{56}
  + a \d_\psi + b\d_\varphi$; and
\item[(b)] $\xi = N_{+2} + \theta_2 R_{34} + \theta_3 R_{56} + a
  \d_\psi + b \d_\varphi$,
\end{enumerate}
and these require further attention.  We can easily discard case (b)
by an argument similar to that employed for the reduction of the M2
brane which we labelled (C) in \cite[Section~3.1.1]{FigSimBranes}.
Although the action is free, there are two types of orbits: at those
points where the Killing vector corresponding to the null rotation
vanishes, the orbits are circles, whereas at other points they are
real lines.  This means that what we have is an action of $\RR$ with
nontrivial stabilisers at those points where the orbits are closed.
As a result the quotient is not smooth.

The vector field $\xi$ in (a) is a vector field on $\RR^6 \times S^3
\subset \RR^{10}$.  It is convenient to identify $\RR^{10}$ with
$\CC^5$ and introduce complex coordinates $(w_1,w_2,w_3,z_1,z_2)$
where $z_1$ and $z_2$ are given by equation \eqref{eq:C2}.  The
integrated action of $\xi$ on $\CC^5$ is given by
\begin{equation*}
  (w_1,w_2,w_3,z_1,z_2) \mapsto \left( e^{i\theta_1 t} w_1,
  e^{i\theta_2 t} w_2, e^{i\theta_3 t} w_3,  e^{i(a+b)t/2}
  z_1, e^{i(a-b)t/2} z_2 \right)~.
\end{equation*}
Generically this defines a curve in the $5$-tori defined by fixing the
values of $|w_i|$ and $|z_i|$.  Away from the nut we must have that at
least one of $z_i$ is nonzero, whereas the $w_i$ are allowed to
vanish.  Unless the integral curves are periodic, the quotient of such
a torus will not be Hausdorff: this is essentially the same situation
as the more familiar irrational flows on the $2$-torus.  To avoid this
situation, there must be some $T>0$ for which the phase factors
\begin{equation*}
  e^{i\theta_i T}~,\qquad e^{i(a+b)T/2} \qquad \text{and} \qquad
  e^{i(a-b)T/2}
\end{equation*}
must all be equal to $1$; equivalently,
\begin{equation*}
  \theta_i T = 2\pi p_i~, \qquad a T = 2\pi (n+m) \qquad \text{and}
  \qquad bT = 2\pi (n-m)~,
\end{equation*}
for some integers $p_i,m,n$.  Taking $T$ to be the period (the
smallest positive number with this property), we must have that
$\gcd(p_1,p_2,p_3, n+m, n-m) = 1$.  In any case, the ratio of any two
of $\theta_i, a, b$ is rational whenever it is defined.

By considering the points where $w_i=0$, we notice that in order to
have a free action, $a$ and $b$ cannot both vanish.  Therefore we must
distinguish between two cases: $a\neq 0$ and $a=0$ (whence $b\neq 0$).
In the first case, $n\neq -m$, so we can rewrite the Killing vector as
\begin{equation*}
  \xi = \frac{a}{n+m} \left( p_1 R_{12} + p_2 R_{34} + p_3 R_{56} +
  (n+m) \d_\psi + (n-m) \d_\varphi \right)~.
\end{equation*}
Using the freedom to rescale $\xi$, we arrive at
\begin{equation*}
  \xi = p_1 R_{12} + p_2 R_{34} + p_3 R_{56} + (n+m) \d_\psi + (n-m)
  \d_\varphi~,
\end{equation*}
where $\gcd(p_1,p_2,p_3,n+m,n-m) = 1$.  Its integral curves are given
by
\begin{equation*}
  \left( e^{ip_1 t} w_1,
    e^{ip_2 t} w_2, e^{ip_3 t} w_3,  e^{int} z_1, e^{imt} z_2
  \right)~,
\end{equation*}
which have period $2\pi$.  If $a=0$, so that $m=-n$, we can rewrite
the vector field as
\begin{equation*}
  \xi = \frac{b}{2n} \left( p_1 R_{12} + p_2 R_{34} + p_3 R_{56} +
  2n \d_\varphi \right)~.
\end{equation*}
We can again rescale to arrive at
\begin{equation*}
  \xi = p_1 R_{12} + p_2 R_{34} + p_3 R_{56} + 2n \d_\varphi~,
\end{equation*}
where $\gcd(p_1,p_2,p_3,2n) = 1$, which integrates to
\begin{equation*}
  \left( e^{ip_1 t} w_1,
    e^{ip_2 t} w_2, e^{ip_3 t} w_3,  e^{int} z_1, e^{-int} z_2
  \right)~,
\end{equation*}
which again has period $2\pi$.  Notice that both cases reduce to
studying the orbits
\begin{equation*}
  \left( e^{ip_1 t} w_1,
    e^{ip_2 t} w_2, e^{ip_3 t} w_3,  e^{int} z_1, e^{imt} z_2
  \right)~,
\end{equation*}
where $\gcd(p_1,p_2,p_3,n-m,n+m) = 1$, where $n$ and $m$ need not be
different, but cannot both be zero.

To obtain a smooth quotient the stabilisers of all points (away from
the nut) must be trivial.  This means that when $z_1$ and $z_2$ are
not both zero, the only solution to
\begin{equation*}
  (w_1,w_2,w_3,z_1,z_2) = \left( e^{ip_1 t} w_1,
    e^{ip_2 t} w_2, e^{ip_3 t} w_3,  e^{int} z_1, e^{imt} z_2
    \right)~,
\end{equation*}
must be $t \in 2\pi\ZZ$.  Considering the points $(0,0,0,0,1)$ and
$(0,0,0,1,0)$ we see that this is the case if and only if $n=\pm 1$
and $m=\pm 1$, giving four cases in total: the two cases where $n=m$
correspond to $b=0$ and the other two cases correspond to $a=0$.
Changing the sign of $\xi$, if necessary, which is the only rescaling
freedom left, we can choose $n=1$.  This gives two cases:
\begin{enumerate}
\item[(i)] $\xi = 2 \d_\psi + p_1 R_{12} + p_2 R_{34} + p_3 R_{56}$,
  and
\item[(ii)] $\xi = 2 \d_\varphi + p_1 R_{12} + p_2 R_{34} + p_3
  R_{56}$,
\end{enumerate}
where $p_i\in\ZZ$ and $\gcd(2,p_1,p_2,p_3) = 1$.  It is easy to see
that there are no further conditions on the $p_i$.

In summary, there are two possible cases of freely-acting spacelike
Killing vectors without translations and hence with integer moduli
$p_i$:
\begin{enumerate}
\item[(D)] $\xi = 2 \d_\psi + p_1 R_{12} + p_2 R_{34} + p_3 R_{56}$,
  with $\gcd(2,p_1,p_2,p_3) = 1$;
\item[(E)] $\xi = 2 \d_\varphi + p_1 R_{12} + p_2 R_{34} + p_3
  R_{56}$, with $\gcd(2,p_1,p_2,p_3) = 1$;
\end{enumerate}
Notice that these cases define genuine circle actions, without the
need to further identify points in spacetime.

\subsection{Supersymmetry}
\label{sec:susymkk}

The Killing spinors in a purely gravitational background are precisely
the parallel spinors with respect to the spin connection.  For the
Kaluza--Klein monopole these are tensor products of parallel spinors
in Minkowski spacetime with parallel spinors in the Taub--NUT space.
In the flat coordinates for Minkowski spacetime, parallel spinors are
simply constant spinors in the half-spin representation of
$\Spin(1,6)$.  The parallel spinors in the Taub--NUT space are
computed explicitly in Appendix~\ref{sec:PSTN}.  The result is that
parallel spinors are in one-to-one correspondence with
positive-chirality spinors for $\Spin(4)$.  Moreover this
correspondence is equivariant with respect to the action of (the spin
cover of) the isometry group: $\SU(2) \times \U(1)$.  This allows us
to easily determine the constraints imposed by supersymmetry on the
reductions classified in the previous section.  As usual, translations
act trivially on spinors and a null rotation simply halves the number
of invariant spinors, so it is only the rotational component which is
constrained.

As explained in Appendix~\ref{sec:groups}, the Killing spinors of the
Kaluza--Klein monopole are in one-to-one correspondence with the
subspace of the eleven-dimensional spinor representation $S_{11}$
which, under $\Spin(1,6) \times \SU(2) \times \U(1)$, transforms as
$[S_7 \otimes S_3]$, where $S_7$ is the unique irreducible spinor
representation of $\Spin(1,6)$, $S_3$ is the fundamental of
$\SU(2)$ and $\U(1)$ acts trivially.  This $\U(1)$ is generated by the
Hopf ``translation'' $\d_\psi$, although one should keep in mind that
from the flat space point of view, this is a self-dual rotation and
certainly acts on spinors in that way.  It is the fact that the
spinors have positive chirality which makes $\d_\psi$ act trivially on
them.

Let us first classify the supersymmetric reductions with translations,
and hence with continuous moduli.  Let $\xi$ be a freely-acting
spacelike Killing vector with rotational component
\begin{equation*}
  \rho = \theta_1 R_{12} + \theta_2 R_{34} + \theta_3 R_{56} + b
  \d_\varphi~,
\end{equation*}
where $\theta_i$ and $b$ are real numbers.  We have not included the
rotation $a \d_\psi$ in $\rho$ since this acts trivially on spinors.
The action of $\rho$ on $[S_7 \otimes S_3]$ can be read off from the
weight decomposition \eqref{eq:weightsMKK}.\footnote{In the notation
  of Appendix~\ref{sec:groups}, $\d_\varphi$ can be thought of as
  $R_{78}$.}  Supersymmetry will be preserved if and only if
\begin{equation*}
  \mu_1 \theta_1 + \mu_2 \theta_2 + \mu_3 \theta_3 + \mu_4 b = 0~,
\end{equation*}
where $\mu_i^2 = 1$.  These equations define a collection of four
hyperplanes in the four-dimensional space parametrised by
$(\theta_i,b)$.  If $\rho$ belongs to precisely one of these
hyperplanes, it is annihilated by precisely two weights in $[S_7
\otimes S_3]$.  If $\rho$ belongs to the intersection of precisely two
hyperplanes, then it is annihilated by four weights.  If in the
intersection of precisely three hyperplanes, then it is annihilated by
eight weights.  Finally the only point in the intersection of more
than three such hyperplanes is the origin which is annihilated by all
sixteen weights.

After these preliminary observations it is easy to read off the
different strata of the ``continuous'' moduli space of supersymmetric
reductions of the Kaluza--Klein monopole.  Using the same nomenclature
for the possible Killing vectors, we find the following cases, which
are summarised in Table~\ref{tab:MKKsusycont}.
\begin{enumerate}
\item[(A)] In this case the rotation $\rho$ has $\theta_1 = 0$ and
  hence must belong to the intersection of an $\fsu(3)$ and an
  $\fso(4) \times \fsu(2)$ subalgebras of $\fso(8)$.  The resulting
  three-dimensional stratum has $\nu=\frac18$.  There is supersymmetry
  enhancement to $\nu=\frac14$ in the two-dimensional locus
  corresponding to rotations $\rho$ which belong to the intersection
  of an $\fsu(2)$ and an $\fso(4) \times \fsu(2)$ subalgebras of
  $\fso(8)$.  In addition there is a one-dimensional locus,
  corresponding to vanishing $\rho$, where the supersymmetry is
  enhanced to $\nu=\frac12$.
\item[(B)] This is the same as (A).
\item[(C)] In this case we have $\theta_1=\theta_2=0$.  We therefore
  have a two-dimensional stratum of supersymmetric reductions with
  $\nu=\frac18$ and consisting of rotations $\rho$ in the intersection
  of an $\fsu(2)$ and an $\fso(4) \times \fsu(2)$ subalgebras of
  $\fso(6)$.  There is a one-dimensional sublocus consisting of
  vanishing $\rho$, where supersymmetry is enhanced to $\nu=\frac14$.
\end{enumerate}

\begin{table}[h!]
  \begin{center}
    \setlength{\extrarowheight}{5pt}
    \begin{tabular}{|>{$}c<{$}|>{$}c<{$}|>{$}c<{$}|>{$}c<{$}|}
      \hline
      \text{Translation} & \text{Subalgebra} & \nu  & \dim\\
      \hline
      \hline
      & \cap \fsu(3) & \frac18 & 3\\
      \text{$\d_+$ or $\d_y$} & \cap \fsu(2) & \frac14~\left( \frac18
      \right) & 2~(2)\\ 
      & \{0\} & \frac12~\left( \frac14 \right) & 1~(1)\\[3pt]
      \hline
    \end{tabular}
    \vspace{8pt}
    \caption{Supersymmetric reductions of the Kaluza--Klein monopole
      with continuous moduli.  The notation $\cap \fh$ in the
      ``Subalgebra'' column indicates that the subalgebra is the
      intersection of $\fh$ with the rotational subalgebra of the
      isometry algebra.  A $\d_y$ means a spacelike translation along
      a Minkowski direction.  The numbers in parentheses indicate
      the values in the presence of a null rotation, which can only
      happen when the translation is spacelike.} 
    \label{tab:MKKsusycont}
  \end{center}
\end{table}

We now move on to discuss the supersymmetric reductions without
translations, and hence with only discrete moduli.  In this case, as
the action is that of a circle, which is not simply-connected, there
exists the question of whether the quotient admits a spin structure.
As shown in Appendix~\ref{sec:spin} for the reductions labelled (D)
and (E), the quotient admits a spin structure if and only if the sum
of the $p_i$ is even.  We will therefore assume from now on that this
is the case.  Notice that since the $p_i$ cannot all be even,
precisely two of them must be odd.

Instead of hyperplanes in the continuous moduli space, supersymmetry
now imposes linear diophantine equations on the integer moduli of the
reductions.  A similar analysis to the one before, but paying
attention to the fact that the $p_i$ are integers which cannot all be
even, yields the following results, which are summarised in
Table~\ref{tab:MKKsusydisc}.
\begin{enumerate}
\item[(D)] In this case the rotation is $\rho = p_1 R_{12} + p_2
  R_{34} + p_3 R_{56}$.  Supersymmetry imposes the linear diophantine
  equation
  \begin{equation*}
    p_1 \mu_1 + p_2 \mu_2 + p_3 \mu_3 = 0~.
  \end{equation*}
  There are clearly an infinite number of solutions to this equation
  for which precisely two of the $p_i$ are odd.  The rotation is
  contained in an $\fsu(3)$ subalgebra of $\fso(6)$ and the reduction
  preserves a fraction $\nu = \frac18$ of the supersymmetry.  There is
  supersymmetry enhancement to $\nu=\frac14$ whenever precisely one of
  the $p_i$ vanishes.  The nonzero two $p_i$ must then be odd
  integers.  Clearly there are again an infinite number of such
  solutions.  In addition there is a unique reduction with
  $\nu=\frac12$ corresponding to (a stack of) D6-branes.
\item[(E)] Here the rotation $\rho$ takes the form $\rho = p_1 R_{12}
  + p_2 R_{34} + p_3 R_{56} + 2 \d_\varphi$.  Supersymmetry imposes
    the equation (after some relabelling)
  \begin{equation*}
    p_1 \mu_1 + p_2 \mu_2 + p_3 \mu_3 = 2~.
  \end{equation*}
  There are clearly an infinite number of solutions with precisely two
  of the $p_i$ odd.  The generic reduction preserves a fraction
  $\nu=\frac1{16}$ of the supersymmetry.  The rotation belongs to the
  intersection of an $\fsu(4)$ and an $\fso(6) \times \fsu(2)$
  subalgebras of $\fso(10)$.  There is supersymmetry enhancement to
  $\nu=\frac18$ in either of two situations: when precisely one of the
  $p_i$ vanishes, which corresponds to the intersection of an
  $\fsu(3)$ subalgebra with an $\fso(6) \times \fsu(2)$ subalgebra of
  $\fso(10)$; and when the equation decouples into two equations: $p_1
  = \pm p_2$ and $p_3 = \pm 2$, say, with $p_1$ and $p_2$ odd.  This
  corresponds to intersecting an $\fsp(1) \times \fsp(1)$ subalgebra.
  One might expect further supersymmetry enhancement by intersecting
  an $\fsu(2)$ subalgebra, but this would require two of the $p_i$ to
  vanish and then the remaining nonzero $p = \pm 2$ would not be odd.
  Therefore no further enhancement takes place.
\end{enumerate}

Although it appears from the above discussion that the D6-brane is
isolated, it actually lives in the same moduli space as case (A) or
(B) with vanishing $\rho$.  In fact, those reductions are all special
cases of reductions by a linear system $a \d_\psi + b \d_1 + c \d_+$
of Killing vectors.  The D6-brane corresponds to $b=c=0$, but one can
clearly deform it without sacrificing supersymmetry by turning on a
minkowskian translation which can be either spacelike or null.

\begin{table}[h!]
  \begin{center}
    \setlength{\extrarowheight}{5pt}
    \begin{tabular}{|>{$}c<{$}|>{$}c<{$}|>{$}c<{$}|}
      \hline
      \rho_{\TN} & \text{Subalgebra} & \nu \\
      \hline
      \hline
      & \fsu(3) & \frac18 \\
      2\d_\psi & \fsu(2) & \frac14 \\
      & \{0\} & \frac12 \\[3pt]
      \hline
      & \cap \fsu(4) & \frac1{16}\\
      2\d_\varphi & \cap \fsp(1)  \times \fsp(1) & \frac18\\
      & \cap \fsu(3) & \frac18\\[3pt]
      \hline
    \end{tabular}
    \vspace{8pt}
    \caption{Supersymmetric reductions of the Kaluza--Klein monopole
      without continuous moduli.  The notation $\cap \fh$ in the
      ``Subalgebra'' column indicates that the subalgebra is the
      intersection of $\fh$ with the rotational subalgebra of the
      isometry algebra.}
    \label{tab:MKKsusydisc}
  \end{center}
\end{table}

\subsection{Explicit reductions}

We shall start by studying the reductions denoted (B) and (C) in
Section~\ref{sec:susymkk}.  We should mention that some of the
configurations described in this section have some overlap with the
content of the paper \cite{Uranga}.  The Killing vector can be written
as $\xi = \partial_z + \lambda$, where $z$ stands for a longitudinal
direction, e.g, $y^1$, and $\lambda$ stands for the infinitesimal
transformation
\begin{equation*}
  \begin{aligned}[m]
    \lambda &= \beta (y^0\partial_3 + y^3\partial_0) + \theta_2
    (y^3\partial_4 - y^4\partial_3) + \theta_3
    (y^5\partial_6-y^6\partial_5) \\
    & {} + a\partial_\psi + b\partial_\varphi ~.
  \end{aligned}
\end{equation*} 

The constant matrix $B$ defined in \eqref{eq:data} is a $7\times 7$
matrix which can be written as
\begin{equation}
  \label{eq:Bmatrixkka}
  B= 
  \begin{pmatrix}
    0 & \beta& 0 & 0 & 0 & 0 & 0 \\
    \beta & 0 & -\theta_2 & 0 & 0 & 0 & 0 \\
    0 & \theta_2 & 0 & 0 & 0 & 0 & 0 \\
    0 & 0 & 0 & 0 & -\theta_3 & 0 & 0 \\
    0 & 0 & 0 & \theta_3 & 0 & 0 & 0 \\
    0 & 0 & 0 & 0 & 0 & 0 & 0 \\
    0 & 0 & 0 & 0 & 0 & 0 & 0 \\
  \end{pmatrix}~,
\end{equation}
in the basis of adapted coordinates \eqref{eq:adapted}
$\{x^0,x^3,x^4,x^5,x^6,\psi,\varphi\}$.  Notice that besides $z$, it
does not act on $\{x^2,r,\theta\}$.  Since $\lambda$ involves
$a\partial_\psi + b\partial_\varphi$, there is a nontrivial $\bC$
vector, which in the same basis used for \eqref{eq:Bmatrixkka} is
written as
\begin{equation*}
  (\bC)^t = (\vec{0},a,b)~.
\end{equation*}

Since the starting configuration is a purely gravitational background,
the type IIA configuration involves a ten dimensional metric, dilaton
and a RR 1-form potential, all other fields vanishing.  We shall
introduce a new notation for the RR 1-form potential $C_1$, to avoid
any confusion with the 1-form potential describing the Dirac monopole
in \eqref{eq:ATN}.  The full configuration looks like
\begin{equation}
  \label{eq:kkreda}
  \begin{aligned}[m]
    g &= \Lambda^{1/2}\left\{ds^2(\EE^{1,5}) + g_{\text{TN}}\right\} -
    \Lambda^{3/2} \left(C_1\right)^2  \\
    C_1 &= \Lambda^{-1}\left\{(B\bx)^i(d\bx)_i +
      bV(r)r^2\,\sin^2\theta d\varphi \right. \\
    & \left. {} + Q^2\,V^{-1}(r)(a-b\cos\theta)(d\psi
      -\cos\theta\,d\varphi)
    \right\} \\
    \Phi &=\tfrac34\log\Lambda ~,
  \end{aligned}
\end{equation}
where $(d\bx)_i = \delta_{ij}d\bx^j$. It depends on an scalar function
$\Lambda$ which is defined as
\begin{equation*}
  \Lambda = 1 + (B\bx)^i(B\bx)_i + b^2\,V(r)r^2\,\sin^2\theta
    + Q^2\,V^{-1}(r)(a-b\cos\theta)^2 ~.
\end{equation*}
Notice that whenever there are fluxbranes being described in ten
dimensions, $\theta_i\neq 0$, $b\neq 0$, there are regions in
spacetime where the string coupling constant becomes strong, and so
the type IIA supergravity description is no longer reliable.

As discussed in Appendix~\ref{sec:action}, the Killing vector
$\partial_\varphi$ acts as a rotation inside $\fsu(2)$, and as such it
should be treated at the same level as the other generators of
rotations in the flat directions along the monopole. Having remarked
this point, the interpretation of the different points in the moduli
space can be given as follows. If all parameters are set to zero,
\eqref{eq:kkreda} describes a ten dimensional Kaluza--Klein monopole.
Whenever $a\neq 0$, and following the same arguments used in
\cite{FigSimBranes}, the interpretation is that of a bound state of a
Kaluza-Klein monopole and D6-branes \cite{CGBoundStates}. Before
discussing the other possibilities, it is interesting to analyse this
bound state closer.  From the parametrisation \eqref{eq:C2}, it is
evident that locally $\partial_\psi\propto R_{78} + R_{9\natural}$.
Thus, the reduction looks very similar to the one defining a flux
5-brane \cite{GSflux}.  Actually both preserve the same
supersymmetries, and by considering the limit $r\to 0$ in
\eqref{eq:kkreda}, one is left with the metric and RR 1-form
\begin{equation*}
  \begin{aligned}[m]
    g &= \Lambda^{1/2}\left\{ds^2(\EE^{1,5}) + \frac{Q}{r}\left((dr)^2
        + r^2((d\theta)^2 + \sin^2\theta
        (d\varphi)^2)\right)\right\}\\
    & {} + \Lambda^{-1/2}\,r\cdot Q(d\psi - \cos\theta d\varphi)^2 \\
    C_1 &= a\,r\cdot Q\Lambda^{-1}(d\psi -\cos\theta d\varphi) ~,
  \end{aligned}
\end{equation*}
depending on the scalar function $\Lambda = 1 + a^2\,r\cdot Q$. By
changing the radial coordinate, $\hat{r} = 2\sqrt{Q\cdot r}$, one
recovers the flux 5-brane first introduced in \cite{GSflux} and
identifies the arbitrary parameter $a$ in the above construction, with
the parameter $\beta$ characterising the fluxbrane ``charge'' through
the relation
\begin{equation*}
  \beta = \pm \frac{a}{2}~.
\end{equation*}
One can thus conclude that the flux 5-brane describes the local
region (close to the branes) of a bound state of Kaluza-Klein
monopoles and D6-branes.

By switching on the remaining parameters, we are adding the
corresponding fluxbranes and nullbranes to the configuration. Thus
whenever $|\beta|= |\theta_2|$, there will be a nullbrane. In that
case, we still have the possibility to add a flux 5-brane (F5-brane),
whenever $\theta_3 \pm b=0$.  Setting $\beta=0$, there are three
different fluxbranes that one can construct:
\begin{enumerate}
\item A flux 3-brane (F3-brane) when $\theta_1 \pm \theta_2 \pm b=0$.
\item A F5-brane when $\theta_2 \pm \theta_3 =0$.
\item Due to the isometries of the background there is a further
  F5-brane when $\theta_1 \pm b = 0$.
\end{enumerate}
All other possibilities would break supersymmetry, and as such, they
can be interpreted in terms of F7-branes (or intersections thereof
with no supersymmetry enhancement) or as quotients by the orbits of
boosts \cite{KOSST, SeibergBC, CorCos, Nekrasov}.  In the
latter case, there would be regions of spacetime with closed timelike
curves.  The supersymmetric configurations are summarised in
Tables~\ref{tab:kk1} and \ref{tab:kk2}.  We use the notation $Fp'$ to
denote those fluxbranes which involve the rotation in the 3-sphere
foliating the original Taub--NUT space.

\begin{table}[h!]
  \begin{center}
    \setlength{\extrarowheight}{3pt}
    \begin{tabular}{|>{$}c<{$}|c|>{$}l<{$}|}
      \hline
      \nu & Object & \multicolumn{1}{c|}{Subalgebra}\\
      \hline
      \hline
      \frac14 & KK+N & \RR \\
      \frac14 & KK$\perp$F5 & \cap \fsu(2) \\
      \frac14 & KK$\perp$F5$^\prime$ & \cap \fsu(2)\\
      \frac{1}{8} & KK+N+F5$^\prime$ & \RR\times\cap \fsu(2) \\
      \frac{1}{8} & KK+F3$^\prime$ & \cap \fsu(3) \\
      \hline
    \end{tabular}
    \vspace{8pt}
    \caption{Supersymmetric configurations of Kaluza--Klein monopoles
      (KK) and fluxbranes and nullbranes}
    \label{tab:kk1}
  \end{center}
\end{table}

\begin{table}[h!]
  \begin{center}
    \setlength{\extrarowheight}{3pt}
    \begin{tabular}{|>{$}c<{$}|c|>{$}l<{$}|}
      \hline
      \nu & Object & \multicolumn{1}{c|}{Subalgebra}\\
      \hline
      \hline
      \frac14 & (KK-D6)+N & \RR \\
      \frac14 & (KK-D6)$\perp$F5 & \cap \fsu(2) \\
      \frac14 & (KK-D6)$\perp$F5$^\prime$ & \cap \fsu(2)\\
      \frac{1}{8} & (KK-D6)+N+F5$^\prime$ & \RR\times\cap \fsu(2) \\
      \frac{1}{8} & (KK-D6)+F3$^\prime$ & \cap \fsu(3) \\[3pt]
      \hline
    \end{tabular}
    \vspace{8pt}
    \caption{Supersymmetric configurations of bound states of a
    Kaluza--Klein monopole and D6-branes (KK-D6) and
    fluxbranes and nullbranes}
    \label{tab:kk2}
  \end{center}
\end{table}

Let us now move to study the reductions referred to as (A) in
Section~\ref{sec:susymkk}. The Killing vector can be written as $\xi =
\partial_z + \lambda$, where $z$ stands for a lightlike direction,
i.e. $y^+$, and $\lambda$ stands for the infinitesimal transformation
\begin{equation*}
    \lambda = \theta_2 (y^3\partial_4-y^4\partial_3) 
    + \theta_3 (y^5\partial_6-y^6\partial_5) + a\partial_\psi +
    b\partial_\varphi ~.
\end{equation*} 

The constant matrix $B$ in \eqref{eq:data} is a $6\times 6$ matrix
which can be written as
\begin{equation}
  \label{eq:Bmatrixkkb}
  B= 
  \begin{pmatrix}
    0 & -\theta_2 & 0 & 0 & 0 & 0 \\
    \theta_2 & 0 & 0 & 0 & 0 & 0 \\
    0 & 0 & 0 & -\theta_3 & 0 & 0 \\
    0 & 0 & \theta_3 & 0 & 0 & 0 \\
    0 & 0 & 0 & 0 & 0 & 0 \\
    0 & 0 & 0 & 0 & 0 & 0 \\
  \end{pmatrix}~,
\end{equation}
in the basis $\{x^3,x^4,x^5,x^6,\psi,\varphi\}$. Notice that besides
$z$, it does not act on $\{x^-,x^2,r,\theta\}$. There is again a
nontrivial $\bC$ vector, which in the same basis used for
\eqref{eq:Bmatrixkkb} is written as
\begin{equation*}
  (\bC)^t = (\vec{0},a,b)~.
\end{equation*}

Adopting the same notation as before, the full type IIA configuration
is written as follows
\begin{equation*}
  \begin{aligned}[m]
    g &= \Lambda^{1/2}\left\{ds^2(\EE^{5}) + g_{\text{TN}}\right\} -
    \Lambda^{3/2} \left(C_1\right)^2  \\
    C_1 &= \Lambda^{-1}\left\{dx^- + (B\bx)^i(d\bx)_i +
      bV(r)r^2\,\sin^2\theta d\varphi \right. \\
    & \left. {} + Q^2\,V^{-1}(r)(a-b\cos\theta)(d\psi
      -\cos\theta\,d\varphi) \right\} \\
    \Phi &=\tfrac34\log\Lambda ~.
  \end{aligned}
\end{equation*}
The configuration depends on a scalar function $\Lambda$ which is
defined as
\begin{equation*}
    \Lambda = (B\bx)^i(B\bx)_i + b^2\,V(r)r^2\,\sin^2\theta
    + Q^2\,V^{-1}(r)(a-b\cos\theta)^2 ~.
\end{equation*}

The physical interpretation of these configurations is not clear to
us. It is straightforward to derive the set of supersymmetric
configurations. Either with ($a\neq 0$) or without D6-branes (a=0), we
can add a F3-brane$^\prime$ when $\theta_2 \pm \theta_3 \pm b=$ and a
F5-brane$^\prime$ when $\theta_2 \pm b=0$. Since $a\neq \pm b$, when
$a\neq 0$, we can also add a standard F5-brane for $\theta_2 \pm
\theta_3=0$.

Finally, we shall analyse the reductions involving discrete moduli.
To begin with, we shall discuss the case referred to as (D) in
Section~\ref{sec:susymkk}.  The Killing vector can be written as $\xi
= \partial_z + \lambda$, where $z$ stands for the compact coordinate
along the Hopf fibre, i.e. $\psi$, and $\lambda$ stands for the
infinitesimal transformation
\begin{equation*}
  \lambda = p_1 (y^1\partial_2 - y^2\partial_1) +
  p_2 (y^3\partial_4-y^4\partial_3) +
  p_3 (y^5\partial_6-y^6\partial_5)~.
\end{equation*} 

The constant matrix $B$ in \eqref{eq:data} is a $6\times 6$ matrix 
\begin{equation}
  \label{eq:Bmatrixkkc}
  B= 
  \begin{pmatrix}
    0 & -p_1 & 0 & 0 & 0 & 0 \\
    p_1 & 0 & 0 & 0 & 0 & 0 \\
    0 & 0 & 0 & -p_2 & 0 & 0 \\
    0 & 0 & p_2 & 0 & 0 & 0 \\
    0 & 0 & 0 & 0 & 0 & -p_3 \\
    0 & 0 & 0 & 0 & p_3 & 0
  \end{pmatrix}~,
\end{equation}
in the basis $\{x^1,\dots,x^6\}$.  Notice that it does not act on the
Taub--NUT space.

After the Kaluza--Klein reduction, the full type IIA configuration is 
written as follows
\begin{equation*}
  \begin{aligned}[m]
    g &= \Lambda^{1/2}\left\{ds^2(\EE^{1,6}) + V(r)ds^2(\EE^3) +
    Q^2\,V^{-1}(r)\cos^2\theta \,(d\varphi)^2\right\} \\
    & -\Lambda^{3/2} \left(C_1\right)^2  \\
    C_1 &= \Lambda^{-1}\left\{(B\bx)^i(d\bx)_i 
    - 2Q^2\,V^{-1}(r)\cos\theta\,d\varphi \right\} \\
    \Phi &=\frac34\log\Lambda~.
  \end{aligned}
\end{equation*}
The configuration depends on an scalar function $\Lambda$ which is
defined as
\begin{equation*}
    \Lambda = (B\bx)^i(B\bx)_i + 4Q^2\,V^{-1}(r)~.
\end{equation*}

It should be clear that when we set all parameters to zero, this gives
raise to the standard half-BPS D6-branes. By switching on different
moduli, one is thus adding fluxbranes.  The new feature, as discussed
before, is that due to the fact that $\partial_\psi$ gives raise to a
circle action, the values of these parameters are no longer
continuous. Table~\ref{tab:D6} summarises the set of supersymmetric
composite configurations of D6-branes and fluxbranes.

\begin{table}[h!]
  \begin{center}
    \setlength{\extrarowheight}{3pt}
    \begin{tabular}{|>{$}c<{$}|c|>{$}l<{$}|}
      \hline
      \nu & Object & \multicolumn{1}{c|}{Subalgebra}\\
      \hline
      \hline
      \frac14 & D6$\perp$F5(2) & \fsu(2) \\
      \frac{1}{8} & D6$\perp$F3(0) & \fsu(3) \\[3pt]
      \hline
    \end{tabular}
    \vspace{8pt}
    \caption{Supersymmetric configurations of D6-branes and
      fluxbranes}
    \label{tab:D6}
  \end{center}
\end{table}

There is one other case referred to as (E) in
Section~\ref{sec:susymkk} involving discrete moduli.  The Killing
vector can be written as $\xi = \partial_z + \lambda$, where $z$
stands for the angular coordinate $\varphi$, and $\lambda$ is the same
as in the previous reduction.  The constant matrix $B$ equals
\eqref{eq:Bmatrixkkc} and the full type IIA configuration is written
as follows
\begin{footnotesize}
  \begin{equation*}
    \begin{aligned}[m]
      g &= \Lambda^{1/2}\left\{ds^2(\EE^{1,6}) + V(r)\left((dr)^2 + r^2
          (d\theta)^2\right) + Q^2\,V^{-1}(r)\,(d\psi)^2\right\} -
      \Lambda^{3/2} \left(C_1\right)^2  \\
      C_1 &= \Lambda^{-1}\left\{(B\bx)^i(d\bx)_i
        - 2Q^2\,V^{-1}(r)\cos\theta\,d\psi \right\} \\
      \Phi &=\tfrac34\log\Lambda ~.
    \end{aligned}
  \end{equation*}
\end{footnotesize}
The configuration depends on an scalar function $\Lambda$ which is
defined as
\begin{equation*}
    \Lambda = (B\bx)^i(B\bx)_i + 4\,V(r)r^2\,\sin^2\theta +
    4Q^2\,V^{-1}(r)\cos^2\theta~.
\end{equation*}

\section*{Acknowledgments}

It is a pleasure to thank Roger Bielawski and David Calderbank for
useful discussions about the Taub--NUT geometry, and Helga Baum,
Felipe Leitner, Andrei Moroianu and Elmer Rees for useful discussions
about spin structures.

We started this work while participating in the programme
\emph{Mathematical Aspects of String Theory} at the Erwin Schrödinger
Institute in Vienna, and it is again a pleasure to thank them for
support and for providing such a stimulating environment in which to
do research.  JMF's participation in this programme was made possible
in part by a travel grant from PPARC.  The work was finished while JMF
was visiting the IHÉS, whom he would like to thank for support.  JS
would like to thank the School of Mathematics of the University of
Edinburgh for hospitality during the final stages of this work.  JMF
is a member of EDGE, Research Training Network HPRN-CT-2000-00101,
supported by The European Human Potential Programme.  The research of
JMF is partially supported by the EPSRC grant GR/R62694/01.  JS is
supported by a Marie Curie Fellowship of the European Community
programme ``Improving the Human Research Potential and the
Socio-Economic Knowledge Base'' under the contract number
HPMF-CT-2000-00480, and in part by a grant from the United
States--Israel Binational Science Foundation (BSF), the European
Research Training Network HPRN-CT-2000-00122 and by Minerva.

Finally, the first author would also like to acknowledge useful
discussions on the subject of this paper with Sonia Stanciu.  These
were some of the last mathematical conversations we enjoyed before the
illness which so tragically cut short her life thrust itself upon us.

\appendix

\section{Group theory and spinors}
\label{sec:groups}

In this appendix we collect some facts about how the spinor
representation of $\Spin(1,10)$ decomposes under certain subgroups.
These results are useful in determining the supersymmetric
Kaluza--Klein reductions of the M-wave and the Kaluza--Klein
monopole.

Let us start by recalling a few facts about the irreducible
representations of $\Spin(1,10)$ and of the Clifford algebra
$\Cl(1,10)$.  The Clifford algebra $\Cl(1,10)$ is isomorphic (as a
real associative algebra) to $\Mat(32,\RR) \oplus \Mat(32,\RR)$, where
$\Mat(n,\RR)$ is the algebra of $n\times n$ real matrices.  This means
that there are two inequivalent irreducible representations: real and
of dimension $32$.  They are distinguished by the action of the centre
which is generated by the volume form
\begin{equation*}
  \dvol(\EE^{1,10}) := \Gamma_{01\cdots\natural}~,
\end{equation*}
which squares to the identity.  Let us assume that a choice has been
made once and for all and let $S_{11}$ denote the corresponding
irreducible representation.  This is an irreducible representation of
$\Spin(1,10)$.

In order to determine which Kaluza--Klein reductions of the M-wave
preserve some supersymmetry, we must decompose the subspace $\ker
\Gamma_+ \subset S_{11}$ into irreducible representations of
$\Spin(9)$ and then determine the weight decomposition in terms of a
Cartan subalgebra of $\fso(9)$.  Let us decompose $S_{11}$ as $S_{11}
= S_{11}^+ \oplus S_{11}^-$, where $S_{11}^\pm = \ker \Gamma_\pm$.
The transverse spin group $\Spin(9)$ acts on $S_{11}$ preserving the
subspaces $S_{11}^\pm$, which are isomorphic under $\Spin(9)$ to the
unique irreducible spinor representation $S_9$.  Cartan subalgebras of
$\fso(9)$ are actually contained in an $\fso(8)$ subalgebra, under
which $S_9$ breaks up 
as $S_9 = S_8^+ \oplus S_8^-$, where now the label $\pm$ refers to the
eight-dimensional chirality.  The weight decomposition of $S_9$
relative to a basis dual to $\{R_{2i-1,2i}\}$ is given by:
\begin{equation}
  \label{eq:weightsMW}
  \text{weights}\left(S_9\right) = \left\{
    (\mu_1,\mu_2,\mu_3,\mu_4) \biggl| \mu_i^2 = 1 \right\}~.
\end{equation}

Finally, we discuss the case of the Kaluza--Klein monopole.  The
relevant subgroup of $\Spin(1,10)$ is now $\Spin(1,6) \times \SU(2)
\times \U(1)$, where $\SU(2) \times \U(1) \subset \Spin(4)$ is the 
``spin cover'' of $\U(2) \subset \SO(4)$.  There is a unique half-spin
representation $S_7$ of $\Spin(1,6)$: it is quaternionic, of complex
dimension $8$.  There are two possible actions of the Clifford algebra
$\Cl(1,6)$ on $S_7$ distinguished by whether $\dvol(\EE^{1,6})$, which
squares to $+\1$, acts as $\pm \1$.  The decomposition
\begin{equation*}
  \dvol(\EE^{1,10}) = \dvol(\EE^{1,6}) \dvol(\EE^4)
\end{equation*}
relates this action to the chirality of the four-dimensional spinor.
Under $\Spin(1,10) \subset \Spin(1,6) \times \Spin(4)$, the
eleven-dimensional spinor representation $S_{11}$ decomposes as
\begin{equation*}
  S = [S_7 \otimes S_4^+] \oplus [S_7 \otimes S_4^-]
\end{equation*}
where $S_4^\pm$ are the positive (resp. negative) chirality half-spin
representations of $\Spin(4) = \SU(2)_- \times \SU(2)_+$, where the
$\mp$ refers to (anti)self-duality.  This implies that $\SU(2)_\pm$
acts trivially on $S_4^\pm$, whereas under $\SU(2)_\mp$, $S_4^\pm$ are
both isomorphic to the fundamental representation $S_3$ of $\SU(2) =
\Spin(3)$, which is quaternionic and of complex dimension $2$.
Therefore the product $S_7 \otimes S_4^\pm$ has a real structure and
as before $[S_7 \otimes S_4^\pm]$ is the underlying real
representation.

As discussed in Appendix~\ref{sec:PSTN}, the Killing spinors of the
Kaluza--Klein monopole are in one-to-one correspondence with spinors
in $S_{11}$ whose four-dimensional chirality is positive.  In other
words, the relevant representation of $\Spin(1,6) \times \SU(2) \times
\U(1)$ is isomorphic to $[S_7 \otimes S_3]$ where the $\U(1)$ factor
acts trivially.  It is easy to determine the weight decomposition of
$[S_7 \otimes S_3]$ under a Cartan subalgebra of $\Spin(6) \times
\SU(2)$.  The general element of such Cartan subalgebra can be written
as
\begin{equation*}
  \theta_1 R_{12} + \theta_2 R_{34} + \theta_3 R_{56} + \theta_4
  R_{78}~.
\end{equation*}
The weights relative to the canonical dual basis to the above
$\{R_{ij}\}$ are given by
\begin{equation}
  \label{eq:weightsMKK}
  \text{weights}\left([S_7\otimes S_3]\right) = \left\{
    (\mu_1,\mu_2,\mu_3,\mu_4) \biggl| \mu_i^2 = 1 \right\}~,
\end{equation}
where the signs are uncorrelated, for a total of $16 =
\dim_\RR[S_7\otimes S_3]$ weights.

\section{Parallel spinors in the Taub--NUT space}
\label{sec:PSTN}

In this section we derive the expression for the parallel spinors in
the Taub--NUT geometry $(M,g_{\TN})$ and exhibit the action of the
isometry group on the parallel spinors.

\subsection{The parallel spinors}
\label{sec:parspin}

Our starting point is the Taub--NUT metric \eqref{eq:metricTN}
written as a cohomogeneity-one space under the action of $\SU(2)$.
In this form, the natural coframe $\{\Theta_m\}$ is given by
\begin{equation*}
  \Theta_1 = V^{1/2} r \sigma_1 \qquad
  \Theta_2 = V^{1/2} r \sigma_2 \qquad
  \Theta_3 = V^{-1/2} Q \sigma_3 \qquad
  \Theta_4 = V^{1/2} dr~.
\end{equation*}
The connection one-forms $\omega_{mn}$, defined by
\begin{equation*}
  d\Theta_m + \omega_{mn} \wedge \Theta_n = 0~,
\end{equation*}
are given by
\begin{equation*}
  \begin{aligned}[m]
    \omega_{12} &= \half (1 - V^{-2} + 2 V^{-1}) \sigma_3\\
    \omega_{13} &= - \half (1 - V^{-1}) \sigma_2\\
    \omega_{14} &= \half (1 + V^{-1}) \sigma_1
    \end{aligned}\qquad
    \begin{aligned}[m]
    \omega_{23} &= \half (1 - V^{-1}) \sigma_1\\
    \omega_{24} &= \half (1 + V^{-1}) \sigma_2\\
    \omega_{34} &= \half (1 - V^{-1})^2 \sigma_3~.    
  \end{aligned}
\end{equation*}
Notice that the anti-self-dual combinations are very simple:
\begin{equation*}
  \omega_{14} + \omega_{23} = \sigma_1~, \qquad
  \omega_{24} - \omega_{13} = \sigma_2 \qquad \text{and} \qquad
  \omega_{12} + \omega_{34} = \sigma_3~,
\end{equation*}
which together with the structure equations \eqref{eq:struceqn} for
the $\sigma_i$, imply that the anti-self-dual components
\begin{equation}
  \label{eq:asdcurv}
  \begin{split}
    \Omega_{12} + \Omega_{34} &= d (\omega_{12} + \omega_{34}) +
    (\omega_{14} + \omega_{23}) \wedge (\omega_{13} - \omega_{24})\\
    \Omega_{14} + \Omega_{23} &= d (\omega_{14} + \omega_{23}) +
    (\omega_{12} + \omega_{34}) \wedge (\omega_{24} - \omega_{13})\\
    \Omega_{24} - \Omega_{13} &= d (\omega_{24} - \omega_{13}) +
    (\omega_{14} + \omega_{23}) \wedge (\omega_{12} + \omega_{34})
  \end{split}
\end{equation}
of the curvature two-form vanish, showing that Taub--NUT is indeed
hyperkähler.

A spinor $\varepsilon$ is parallel if and only if it satisfies
\begin{equation*}
  \nabla \varepsilon := d\varepsilon + \sum_{m<n} \omega_{mn}
  \Sigma_{mn} \varepsilon = 0~,
\end{equation*}
where $\Sigma_{mn} = \half \Gamma_{mn}$ are the spin generators and
$\Gamma_n$ is a basis for the Clifford algebra adapted to the 
coframe $\Theta_n$ and obeying $\Gamma_m \Gamma_n + \Gamma_n \Gamma_m
= 2 \delta_{mn} \1$.

We notice that there is no $dr$ in the expressions for $\omega_{mn}$,
whence $\nabla_r = \d_r$ and hence parallel spinors do not depend on
$r$.  This means that we can compute them for any value of $r$. It is
convenient to compute them in the limit $r\to\infty$ since the
asymptotic geometry is flat.  Let $\bar\omega_{mn}$ denote the
connection one-forms in this limit.  Noticing that $V \to 1$ in this
limit, we easily find
\begin{equation*}
  \begin{aligned}[m]
    \bar\omega_{12} &= \sigma_3\\
    \bar\omega_{13} &= 0\\
    \bar\omega_{14} &= \sigma_1
    \end{aligned}\qquad
    \begin{aligned}[m]
    \bar\omega_{23} &= 0\\
    \bar\omega_{24} &= \sigma_2\\
    \bar\omega_{34} &= 0~.    
  \end{aligned}
\end{equation*}
After a short calculation, the parallel spinors in this limit are
given by
\begin{equation}
  \label{eq:PSTN}
  \varepsilon = \exp(-\psi\Sigma_{12}) \,  \exp(-\theta \Sigma_{14})
  \,   \exp(-\varphi \Sigma_{12})\, \varepsilon_0~,
\end{equation}
where $\varepsilon_0$ is a constant spinor.

We now impose that the spinor $\varepsilon$ given above be indeed
parallel.  For this it is convenient to introduce the tensor $T \in
\Omega^1(M) \otimes \fso(TM)$:
\begin{equation*}
  T_{mn} = \omega_{mn} - \bar\omega_{mn}~,
\end{equation*}
which measures the difference between the connection one-forms of the
Taub--NUT geometry and its flat asymptotic limit.  It is explicitly
given by
\begin{equation*}
  \begin{aligned}[m]
    T_{12} &= -\half (1 - V^{-1})^2 \sigma_3\\
    T_{13} &= -\half (1 - V^{-1}) \sigma_2 \\
    T_{14} &= -\half (1 - V^{-1}) \sigma_1
    \end{aligned}\qquad
    \begin{aligned}[m]
    T_{23} &= \half (1 - V^{-1}) \sigma_1\\
    T_{24} &= -\half (1 - V^{-1}) \sigma_2\\
    T_{34} &= \half (1 - V^{-1})^2 \sigma_3~.
  \end{aligned}
\end{equation*}
For any spinor $\varepsilon$ of the form given in \eqref{eq:PSTN},
\begin{equation*}
  \nabla \varepsilon = \sum_{m<n} T_{mn} \Sigma_{mn}
  \varepsilon~,
\end{equation*}
whence it will be a parallel spinor in the Taub--NUT geometry if and
only if it is annihilated by
\begin{multline*}
  \sum_{m<n} T_{mn} \Sigma_{mn} = \half (1-V^{-1}) \sigma_1
  (\Sigma_{23} - \Sigma_{14})\\
  - \half (1-V^{-1}) \sigma_2 (\Sigma_{13} + \Sigma_{24})
  - \half (1-V^{-1})^2 \sigma_3 (\Sigma_{12} - \Sigma_{34})~.
\end{multline*}
Since the one-forms $\sigma_i$ are linearly independent, this is
equivalent to $\varepsilon$ being annihilated by the self-dual
combinations $\Sigma_{23} - \Sigma_{14}$, $\Sigma_{13} + \Sigma_{24}$
and $\Sigma_{12} - \Sigma_{34}$; but these three equations are
equivalent to the chirality condition $\Gamma_{1234} \varepsilon = -
\varepsilon$.  Notice that the chiralities of $\varepsilon$ and
$\varepsilon_0$ agree, whence this equation is equivalent to
$\Gamma_{1234} \varepsilon_0 = -\varepsilon_0$.  Noticing that the
orientation of Taub--NUT is given by
\begin{equation*}
  \dvol_{\TN} = \Theta_4 \wedge \Theta_1 \wedge \Theta_2 \wedge
  \Theta_3~,
\end{equation*}
we can write the chirality condition more invariantly as
\begin{equation}
  \label{eq:chiral}
  \dvol_{\TN} \varepsilon = \varepsilon~.
\end{equation}

\subsection{The action of the isometry group}
\label{sec:action}

Acting on a Killing spinor $\varepsilon$, the expression for the Lie
derivative along a Killing vector $\xi$ becomes an algebraic
condition.  In the case of a parallel spinor, this simplifies to
\begin{equation*}
  \eL_\xi \varepsilon = \tfrac14 d\xi^\flat\, \varepsilon
\end{equation*}
where $\xi^\flat$ is the one-form dual to $\xi$.  Let $\varepsilon =
\Psi(\psi,\theta,\varphi) \varepsilon_0$ denote a parallel spinor,
where $\Psi(\psi,\theta,\varphi)$ denotes the product of
exponentials in equation \eqref{eq:PSTN}.  Since the action of a
Killing vector $\xi$ preserves the space of parallel spinors, it
follows that
\begin{equation*}
  \eL_\xi \varepsilon = \Psi(\psi,\theta,\varphi) M_\xi
  \varepsilon_0~,
\end{equation*}
for some \emph{constant} endomorphism $M_\xi$ of the chiral spinor
representation.  Since $M_\xi$ is constant, the calculation can be
simplified by choosing a convenient point in which the expressions
simplify.  A straightforward calculation reveals that for $\xi_i$,
$i=1,2,3,4$ given by \eqref{eq:KVTN}, the endomorphisms $M_{\xi_i}$
are given by
\begin{equation*}
  M_{\xi_1} = -\Sigma_{13} \qquad
  M_{\xi_2} = -\Sigma_{23} \qquad
  M_{\xi_3} = -\Sigma_{12} \qquad
  M_{\xi_4} = 0~.
\end{equation*}
One can check that they satisfy the $\fsu(2) \times \fu(1)$ Lie
algebra, as expected; in particular, for $i=1,2,3$,
\begin{equation*}
  [M_{\xi_i}, M_{\xi_j}] = \varepsilon_{ijk} M_{\xi_k}~.
\end{equation*}

In summary, the parallel spinors in the Taub--NUT geometry are given
by \eqref{eq:PSTN}, where $\varepsilon_0$ (and hence $\varepsilon$) is
subject to the chirality condition \eqref{eq:chiral}.  Furthermore
the correspondence is equivariant under the action of $\fsu(2) \times
\fu(1) \subset \fso(4)$, where the $\fu(1)$ is self-dual (and hence
acts trivially on positive-chirality spinors) and $\fsu(2)$ is
anti-self-dual and hence acts on positive-chirality spinors via the
fundamental representation.

\section{Spin structures on quotients}
\label{sec:spin}

Let $M$ be a spin manifold and let $\Gamma$ be a group acting freely
on $M$ via orientation-preserving isometries.  Is the quotient
$M/\Gamma$ spin?  The answer to this question is affirmative if and
only if there is a spin bundle on $M$ to which the action of $G$
lifts.  Let $P_{\SO}(M)$ denote the bundle of oriented orthonormal
frames.  This is a principal $\SO(p,q)$-bundle, where $M$ has
signature $(p,q)$.  Let $\vartheta_0: \Spin(p,q) \to \SO(p,q)$ be the
spin double-cover.  A spin structure on $M$ is a principal
$\Spin(p,q)$-bundle $P_{\Spin}(M)$ and a bundle map $\vartheta:
P_{\Spin}(M) \to P_{\SO}(M)$ which restricts to $\vartheta_0$ on the
fibres.  Since $\Gamma$ acts via orientation-preserving isometries, it
acts on $P_{\SO}(M)$.  The question is whether this action lifts to
$P_{\Spin}(M)$ in such a way that $\vartheta$ is equivariant.  This
seems to be a difficult question to settle in general; but it is
possible to decide for the groups $\Gamma$ of interest acting on the
Kaluza--Klein monopole $M = \RR^{1,6} \times \TN$.

Since $M$ is contractible, the frame bundle is trivial:
\begin{equation*}
  P_{\SO}(M) \cong \SO(1,10) \times M~,
\end{equation*}
where the isomorphism is given explicitly by choosing an oriented
orthonormal frame $\be_a$.  Given any other oriented orthonormal frame
$\be'_a$, they are related as follows:
\begin{equation*}
  \be'_a(x) = \sum_b \be_b(x) L^b{}_a(x)~,
\end{equation*}
where $L: M \to \SO(1,10)$ is a local Lorentz transformation.

As discussed before, the isometry group of $M$ is $(\SO(1,6) \ltimes
\RR^7) \times \U(2)$.  Given $g\in G$, we have a homonymous map $g: M
\to M$ sending $x \mapsto g\cdot x$ and its derivative map $g_* : T_xM
\to T_{g\cdot x}M$.  If $\be_a(x)$ is an orthonormal frame at $x$,
then $g_* \cdot \be_a(x)$ is an orthonormal frame at $g\cdot x$.
Therefore there exists an element $S(g, x) \in \SO(1,10)$ such that
\begin{equation*}
  g_* \cdot \be_a(x) = \sum_b \be_b(g\cdot x) S^b{}_a(g,x)~.
\end{equation*}
In other words, $S$ defines a map $G \times M \to \SO(1,10)$.  By
considering the action of $g_*$ on another frame $\be'_a(x)$, we see
that the action of $G$ on $P_{\SO}(M) \cong \SO(1,10) \times M$
consists of left multiplication on $\SO(1,10)$ and $g$ on $M$:
\begin{equation*}
  g_* \cdot (L(x),x) = (S(g,x) L(x), g\cdot x)~.
\end{equation*}
Since the map $g \mapsto g_*$ is a homomorphism, the map $S:G \times M
\to \SO(1,10)$ satisfies the following identity \footnote{This
  identity says that $S$ is a contravariant functor between the
  grupoids $G \times M$ and $\SO(1,10)$.  We thank Takashi Kimura for
  this observation.} for all $g_1,g_2\in G$ and $x\in M$:
\begin{equation}
  \label{eq:functor}
  S(g_1 g_2, x) = S(g_1, g_2 \cdot x) S(g_2, x)~.
\end{equation}

Now let $\Gamma \subset G$ be a subgroup and consider the restriction
to $\Gamma$ of the map $S:\Gamma \times M \to \Spin(1,10)$.  We are
interested primarily in connected one-parameter subgroups $\Gamma$,
which are thus diffeomorphic to either $\RR$ or $S^1$.  Since $M$ is
simply-connected, there is a unique spin bundle $P_{\Spin}(M)$.  The
action of $\Gamma$ on $P_{\SO}(M)$ will lift to $P_{\Spin}(M)$ if and
only if there exists a lift
\begin{equation*}
  \xymatrix{ & \Spin(1,10) \ar[d]^{\vartheta_0} \\
  \Gamma \times M \ar@{.>}[ur]^? \ar[r]^S & \SO(1,10) }
\end{equation*}
making the above diagram commutative.  If $\Gamma \cong \RR$ then the
lift exists on purely topological grounds, because both $\Gamma \times
M$ and $\Spin(1,10)$ are simply connected.  Therefore we need only
consider the cases where $\Gamma \cong S^1$.

Let us trivialise the frame bundle by choosing a frame.  Equivalently,
we find it more convenient to choose a coframe made out of the
$dx^\mu$ for $\RR^{1,6}$ and the coframe $\Theta_m$ introduced in
Section~\ref{sec:parspin} for Taub--NUT.  The action of the isometry
group on this coframe is given as follows: the translations $\RR^7$
act trivially, whereas $\SO(1,6)$ acts linearly through the
fundamental representation.  The action of $\U(2)$ on the Taub--NUT
coframe is such that $\sigma_1$ and $\sigma_2$ are rotated, whereas
$r$ and $\sigma_3$ are inert.  Therefore on this frame only the
$\U(1)$ generated by $\d_\psi$ acts nontrivially and generates a
circle action on the plane spanned by $(r V^{\half} \sigma_1, r
V^{\half} \sigma_2)$.  In any case, the map $\Gamma \times M \to
\SO(1,10)$ is constant on $M$, and hence equation \eqref{eq:functor}
says that it defines a homomorphism $\Gamma \to \SO(1,10)$.  The
question is whether this homomorphism lifts to $\Spin(1,10)$.  This
condition is equivalent to the existence of a homomorphism $\Gamma \to
\Spin(1,10)$ whose image is mapped \emph{isomorphically} under
$\vartheta_0$ to the image of $\Gamma$ in $\SO(1,10)$.

This question can be analysed as follows.  Since $\Gamma$ is a
one-parameter group, its image $\overline\Gamma \subset \SO(1,10)$ is
generated by an element $X$ of the Lie algebra $\fso(1,10)$.  We will
embed $\fso(1,10)$ in the Clifford algebra $\Cl(1,10)$ and we will
exponentiate $X$ there: the result is a one-parameter group
$\widehat\Gamma \subset \Spin(1,10)$ which covers $\overline\Gamma$.  The
question is then whether $\widehat\Gamma$ is isomorphic to
$\overline\Gamma$ or covers it twice.  It is clear that the latter
will be the case if and only if $-\1 \in \widehat\Gamma$.

Let us investigate this question for the two cases of interest:
\begin{itemize}
\item[(D)] $\xi = 2 \d_\psi + p_1 R_{12} + p_2 R_{34} + p_3 R_{56}$,
  and
\item[(E)] $\xi = 2 \d_\varphi + p_1 R_{12} + p_2 R_{34} + p_3
  R_{56}$.
\end{itemize}
These Killing vectors define elements of the isometry algebra $\fg$
and hence induce elements in $\fso(1,10)$ by the derivative at the
identity of the map $G \to \SO(1,10)$ described above.  Introducing 
the notation $\gamma_{ij}$ for the generators of the Clifford algebra
$\Cl(4)$ corresponding to the Taub--NUT directions, we can write the
elements $X \in \fso(1,10) \subset \Cl(1,10)$ corresponding to the
above Killing vectors as follows
\begin{itemize}
\item[(D)] $X^{(D)} = 2 \gamma_{12} + p_1 \Gamma_{12} + p_2
  \Gamma_{34} + p_3 \Gamma_{56}$, and
\item[(E)] $X^{(E)} = p_1 \Gamma_{12} + p_2 \Gamma_{34} + p_3
  \Gamma_{56}$;
\end{itemize}
where we have used that $\d_\varphi$ acts trivially on the chosen
frame.  Exponentiating the above elements in the Clifford algebra we
obtain subgroups of $\Spin(1,10)$ with elements
\begin{multline*}
  \widehat g^{(D)}(t) := \exp\left(\frac{t}2 X^{(D)}\right) = \left(\1
    \cos t + \gamma_{12} \sin t\right)
  \left(\1 \cos \frac{p_1 t}2 + \Gamma_{12} \sin \frac{p_1 t}2
  \right)\\
  \times  \left(\1 \cos \frac{p_2 t}2 + \Gamma_{34} \sin \frac{p_2 t}2 \right)
  \left(\1 \cos \frac{p_3 t}2 + \Gamma_{56} \sin \frac{p_3 t}2 \right)~,
\end{multline*}
and
\begin{multline*}
  \widehat g^{(E)}(t) := \exp\left(\frac{t}2 X^{(E)}\right) = \left(\1
    \cos \frac{p_1 t}2 + \Gamma_{12} \sin \frac{p_1 t}2 \right)\\
  \times
  \left(\1 \cos \frac{p_2 t}2 + \Gamma_{34} \sin \frac{p_2 t}2 \right)
  \left(\1 \cos \frac{p_3 t}2 + \Gamma_{56} \sin \frac{p_3 t}2
  \right)~,
\end{multline*}
for $t \in [0,4\pi]$.  Each of these groups is either a double-cover
of the image $\overline\Gamma$ of $\Gamma$ in $\SO(1,10)$ or
isomorphic to it.  It will be a double cover if and only if it
contains $-\1$, since this is the only other point in the pre-image
under $\vartheta_0$ of the identity in $\SO(1,10)$.  A moment's
thought reveals that this can only occur for $t=2\pi$ where
\begin{equation*}
  g^{(D)}(2\pi) = g^{(E)}(2\pi) = (-1)^{p_1+p_2+p_3} \1~.
\end{equation*}
Therefore we conclude that this will contain $-\1$ if and only if the
sum of the $p_i$ is odd.  In other words, $M/\Gamma$ is spin if and
only if $p_1 + p_2 + p_3$ is even.

\bibliographystyle{utphys}
\bibliography{AdS,ESYM,Sugra,Geometry,CaliGeo}

\end{document}